\documentclass[twocolumn]{aastex631}

\usepackage[T1]{fontenc}
\usepackage{ae,aecompl}
\usepackage{multirow}
\usepackage{tabularx}
\usepackage{supertabular}
\usepackage{longtable}
\usepackage[]{url}
\newcommand{\ztfink}[1]{ZTF-Fink}

\begin{document}

\title{GRANDMA and HXMT Observations of GRB 221009A - the Standard-Luminosity Afterglow of a Hyper-Luminous Gamma-Ray Burst - In Gedenken an David Alexander Kann}

%\author[0000-0002-0786-7307]{Greg J. Schwarz}

--------
\author[0000-0003-2902-3583]{D.~A.~Kann}\thanks{Corresponding author: grandma-l@in2p3.fr}
\affiliation{Hessian Research Cluster ELEMENTS, Giersch Science Center, Max-von-Laue-Stra{\ss}e 12, Goethe University Frankfurt, Campus Riedberg, 60438 Frankfurt am Main, Germany}
\author[0009-0007-3673-8997]{S.~Agayeva}
\affiliation{N. Tusi Shamakhy Astrophysical Observatory, Azerbaijan National Academy of Sciences, settl. Y. Mammadaliyev, AZ 5626, Shamakhy, Azerbaijan}
\author{V.~Aivazyan}
\affiliation{E. Kharadze Georgian National Astrophysical Observatory, Mt. Kanobili, Abastumani, 0301, Adigeni, Georgia}
\affiliation{Samtskhe-Javakheti State University, Rustaveli Str. 113, Akhaltsikhe, 0080, Georgia}
\author{S.~Alishov}
\affiliation{N. Tusi Shamakhy Astrophysical Observatory, Azerbaijan National Academy of Sciences, settl. Y. Mammadaliyev, AZ 5626, Shamakhy, Azerbaijan}
\author[0009-0004-9687-3275]{C.~M.~Andrade}
\affiliation{School of Physics and Astronomy, University of Minnesota, Minneapolis, Minnesota 55455, USA}
\author[0000-0002-7686-3334]{S.~Antier}
\affiliation{Artemis, Observatoire de la C\^ote d'Azur, Universit\'e C\^ote d'Azur, Boulevard de l'Observatoire, 06304 Nice, France}
\author[0000-0002-9808-1990]{A.~Baransky}
\affiliation{Astronomical Observatory of Taras Shevchenko National University of Kyiv, Observatorna Str. 3, Kyiv, 04053, Ukraine}
\author{P.~Bendjoya}
\affiliation{Laboratoire J.-L. Lagrange, Universit\'e de Nice Sophia-Antipolis, CNRS, Observatoire de la C\^ote d'Azur, 06304 Nice, France}
\author[0000-0001-6285-9847]{Z.~Benkhaldoun}
\affiliation{Oukaimeden Observatory, High Energy Physics and Astrophysics Laboratory, FSSM, Cadi Ayyad University, Av. Prince My Abdellah, BP 2390 Marrakesh, Morocco}
\author[0000-0002-8291-2817]{S.~Beradze}
\affiliation{E. Kharadze Georgian National Astrophysical Observatory, Mt. Kanobili, Abastumani, 0301, Adigeni, Georgia}
\affiliation{Samtskhe-Javakheti State University, Rustaveli Str. 113, Akhaltsikhe, 0080, Georgia}
% ...

\begin{abstract}

GRB 221009A is the brightest Gamma-Ray Burst (GRB) detected in more than 50 years of study. In this paper, we present observations in the X-ray and optical domains ranging from the prompt emission (optical coverage by all-sky cameras) up to 20 days after the GRB obtained by the GRANDMA Collaboration (which includes observations from more than 30 professional and amateur telescopes) and the \textit{Insight}-\textit{HXMT} Collaboration operating the X-ray telescope \textit{HXMT}-LE.
We study the optical afterglow with empirical fitting using the GRANDMA+\textit{HXMT}-LE data sets, augmented with data from the literature up to 60 days.
We then model numerically, using a Bayesian approach, the GRANDMA and \textit{HXMT}-LE afterglow observations, that we augment with \textit{Swift}-XRT and additional optical/NIR observations reported in the literature. We find that the GRB afterglow, extinguished by a large dust column, is most likely behind a combination of a large Milky-Way dust column combined with moderate low-metallicity dust in the host galaxy. 
Using the GRANDMA+\textit{HXMT}-LE+XRT dataset, we find that the simplest model, where the observed afterglow is produced by synchrotron radiation at the forward external shock during the deceleration of a top-hat relativistic jet by a uniform medium, fits the multi-wavelength observations only moderately well, with a tension between the observed temporal and spectral evolution. This tension is confirmed when using the extended dataset. We find that the consideration of a jet structure (Gaussian or power-law), the inclusion of synchrotron self-Compton emission, or the presence of an underlying supernova do not improve the predictions, showing that the modelling of GRB22109A will require going beyond the most standard GRB afterglow model. Placed in the global context of GRB optical afterglows, we find the afterglow of GRB 221009A is luminous but not extraordinarily so, highlighting that some aspects of this GRB do not deviate from the global known sample despite its extreme energetics and the peculiar afterglow evolution.

\end{abstract}


\section{Introduction}

Gamma-ray bursts (GRBs) are among the most energetic phenomena detected in the Universe. They release extreme amounts of energy in soft $\gamma$-rays, up to $1M_\odot$ assuming isotropic emission \citep{1999Natur.398..389K,2017ApJ...837..119A}, and can also be exceedingly luminous in the optical domain \citep{1999Natur.398..400A,2006ApJ...638L..71B,2007AJ....133.1187K,2008Natur.455..183R,2009ApJ...691..723B,2023arXiv230102407J}.

GRBs exhibit durations\footnote{Usually measured as $T_{90}$, denoting the time span during which 90\% of the emission, from 5\% to 95\%, are accumulated. $T_{90}$ durations are detector-dependent and can include $\gamma$-ray tail emission in bright bursts.} from ms up to several hours \citep[e.g.,][]{2011Natur.480...72T,2013ApJ...766...30G,2014ApJ...781...13L}. They have been historically divided \citep{1981Ap&SS..80....3M,1993ApJ...413L.101K} into two classes based on their duration and spectral hardness.

So-called ``short/hard GRBs'' have durations of a few seconds or less and a harder spectrum with respect to their isotropic energy release \citep[e.g.,][]{2020MNRAS.492.1919M,2021arXiv210913838A}. They have been linked to gravitational waves \citep{2017PhRvL.119p1101A,2017ApJ...848L..12A,2017ApJ...848L..13A,2017ApJ...848L..14G}, and their progenitors are supposed to be mainly coalescing compact objects such as binary neutron stars or neutron-star black-hole binary systems \cite[for reviews, see][]{2007PhR...442..166N,2014ARA&A..52...43B}. The general ``short/hard'' paradigm has been called into question especially with the long-duration event GRB 211211A \citep{2022Natur.612..223R,2022Natur.612..228T,2022Natur.612..232Y}, which has been claimed to be associated with kilonova emission, a hallmark of compact binary mergers. 

Conversely, so-called ``long/soft GRBs'' generally have durations greater than a few seconds, a softer spectrum, and their origin is most likely related to the core-collapse of rapidly rotating massive stars \citep{1993ApJ...405..273W,mosta}. Similar to the case of short GRBs, the ``long/soft'' paradigm has been called into question by GRB 200826A, a sub-second GRB clearly associated with supernova (SN) emission \citep{2021NatAs...5..917A,2021NatAs...5..911Z,2022ApJ...932....1R}. For reviews of long GRBs and their connection to stripped-envelope supernova explosions, see \cite{2009ARA&A..47..567G,2012grb..book..169H,2017AdAst2017E...5C}.

The luminosity of GRB afterglows (in the X-ray to optical/Near-InfraRed [NIR] energy range) is moderately correlated with the isotropic prompt-emission (mostly $\gamma$-ray) energy release $E_{\rm iso}$ \citep{2008ApJ...689.1161G,2009ApJ...701..824N,2010ApJ...720.1513K,2011ApJ...734...96K}, so very luminous GRBs usually have more luminous afterglows, and of course a low distance also implies a brighter afterglow that can be more easily followed-up. A combination of these two features therefore usually yields the richest data sets for any electromagnetic study. Two examples of such well-studied, nearby bright GRBs are GRBs 030329 and 130427A. GRB 030329 occurred at $z=0.16867\pm0.00001$ \citep{2007ApJ...671..628T}, and is to this day the GRB afterglow with the most optical/NIR observations. It yielded data for a wide range of studies on the prompt emission, afterglow evolution and polarization, and the associated SN 2003dh \citep{2004ApJ...617.1251V,2004ApJ...606..381L,2003Natur.426..157G,2003Natur.423..847H,2003ApJ...591L..17S,2003ApJ...599..394M}. The second being GRB 130427A, the first known nearby GRB \citep[$z=0.3399\pm0.0002$,][]{2019A&A...623A..92S} that exhibited an $E_{\rm iso}$ in the range of ``cosmological'' GRBs at $z\gtrsim1$. There is also a rich observational data set for this event, stretching from trigger time to nearly 100 Ms \citep[e.g.][]{2014Sci...343...48M,2014Sci...343...38V,2014Sci...343...42A,2014ApJ...781...37P,2014A&A...567A..29M,2014ApJ...792..115L,2014MNRAS.444.3151V,2016MNRAS.462.1111D}.

In this paper, we report observations by the GRANDMA collaboration and its partners of the paragon of nearby, bright GRBs, GRB 221009A, by far the brightest GRB observed to date.

On 9 October 2022, at 14:10:17 UT, the Burst Alert Telescope \citep[BAT,][]{2005SSRv..120..143B} onboard the \textit{Neil Gehrels Swift Observatory} satellite \citep[][\textit{Swift} hereafter]{2004ApJ...611.1005G} triggered and located a new, X-ray bright transient denoted as \textit{Swift} J1913.1+1946 \citep[triggers 1126853 and 1126854,][]{2022ATel15650....1D,2022GCN.32632....1D}. \textit{Swift} slewed immediately to the position and its narrow-field instruments, the X-ray telescope \citep[XRT,][]{2005SSRv..120..165B} and the Ultra-Violet/Optical Telescope \citep[UVOT,][]{2005SSRv..120...95R} discovered a transient, which was very bright in X-rays ($>800$ ct/s) and moderately bright in the optical (unfiltered finding chart, $white=16.63\pm0.14$ mag). The optical detection was somewhat remarkable as the transient lies in the Galactic plane and extinction along the line-of-sight is very high, $E_{(B-V)}=1.32$ mag/$A_V=4.1$ mag \citep[][henceforth SF11]{Schlafly2011}. It was furthermore reported that the source was also detected over ten minutes earlier by the Gas-Slit Camera (GSC) of the \textit{MAXI} X-ray detector onboard the International Space Station \citep[ISS,][]{2022ATel15651....1N,2022GCN.32756....1K,2023Swift}. Overall, this is in agreement with a new Galactic transient.

About 6.5 hours after the \textit{Swift} trigger, it was reported by \cite{2022GCN.32635....1K} that this source may be a GRB, GRB 221009A, as
both the Gamma-Ray Burst Monitor \citep[GBM,][]{2009ApJ...702..791M}
 and the Large Area Telescope \citep[LAT,][]{2009ApJ...697.1071A} of the \textit{Fermi} observatory \citep{1999APh....11..277G} triggered on a GRB\footnote{The initial GBM trigger notice was distributed, but a problem with automated data processing prevented any additional real-time classification/localization messages from being sent to the ground.} localized to the same sky position at 13:16:59.99 UT, which we henceforth use as trigger time $T_0$. This event turned out to be extraordinarily bright \citep{2022GCN.32636....1V}, not just the brightest event ever detected by GBM, but the brightest \textit{ever} detected.

 The event begins with a moderately bright precursor, followed by $\approx180$ s of quiescence before the main phase starts. The first peak, $\approx20$ s long, would already place GRB~221009A among the brightest GRBs ever detected, exceeding all but a handful of GBM/Konus detections. This peak is followed by two ultra-bright peaks, and finally a fourth, less bright but longer peak which fades into a high-energy afterglow at $\approx600$ s. The extreme fluence led to a saturation of all sensitive $\gamma$-ray detectors, such as \textit{Fermi} GBM
\citep{2022GCN.32642....1L}, \textit{Fermi} LAT \citep{2022GCN.32637....1B,2022GCN.32658....1P,2022GCN.32760....1O,2022GCN.32916....1O}, Konus-\textit{Wind} \citep{2022GCN.32668....1F,2023arXiv230213383F}, \textit{Insight}-HXMT/HE \citep{2022ATel15660....1T,2022ATel15703....1G}, \textit{AGILE}/MCAL+AC \citep{2022GCN.32650....1U}, and \textit{INTEGRAL} SPI/ACS \citep{2022GCN.32660....1G}.

 This saturation leads to preliminary analyses reporting only lower limits on the true fluence. \textit{INTEGRAL} SPI/ACS
 \citep{2022GCN.32660....1G}  analysis finds 1.3$\times10^{-2}$ erg/cm$^2$, \textit{Fermi} GBM finds ($2.912\pm0.001$)$\times10^{-2}$ erg/cm$^2$ and peak flux $2385\pm3$ ph s$^{-1}$ cm$^{-2}$ \citep{2022GCN.32642....1L}, Konus-\textit{Wind} report 5.2$\times10^{-2}$ erg/cm$^2$ \citep{2022GCN.32668....1F}, and \cite{2022GCN.32762....1K} estimate $\approx9$ $\times10^{-2}$ erg/cm$^2$. Even these preliminary estimates show GRB 221009A exceeded GRB 130427A in fluence by a factor of at least 10. Recently, \cite{2023arXiv230213383F} presented the full Konus-\textit{Wind} analysis, which yields a fluence more than twice as high as that derived by \cite{2022GCN.32762....1K}, leading to an isotropic energy release $E_{iso}>10^{55}$ erg, twice as high as the previous record holder.

Several smaller orbital detectors were not saturated, stemming from size, environment, or off-axis detection, such as detectors on \textit{Insight} \citep[the Low-Energy (LE) telescope and the Particle Monitors,][]{2022ATel15703....1G},
\textit{SATech-01/GECAM-C} HEBS \citep{2022GCN.32751....1L}, \textit{GRB\-Alpha} \citep{2022GCN.32685....1R,2023arXiv230210047R}, \textit{STPSat-6}/SIRI-2 \citep{2022GCN.32746....1M}, and \textit{SRG}/ART-XC \citep{2022GCN.32663....1L,2023arXiv230213383F}.

 Optical spectroscopy of the transient showed it to indeed be a GRB afterglow, with a redshift $z=0.151$ measured both in absorption and emission \citep{2022GCN.32648....1D,2022GCN.32686....1C,2022GCN.32765....1I,2023arXiv230207891M}, making it even closer than GRB 030329. Such an event is ultra-rare, e.g., \cite{2022GCN.32793....1A} estimate it to occur only once every half-millennium. \cite{2023Swift}, \cite{2023arXiv230207906O} and \cite{2023arXiv230207891M} also discuss the rate of events, finding estimates of the same order. Using the significantly higher fluence from \cite{2023arXiv230213383F}, \cite{2023arXiv230214037B} derive an even more extreme value, finding a repetition rate of 1 every $\approx10,000$ years, a once-in-all-human-civilization event.

The GRB showed very strong VHE emission, with a $\approx400$ GeV photon detected by \textit{Fermi} LAT \citep{2022GCN.32748....1X,2022arXiv221013052X}, a highly significant detection by \textit{AGILE}/GRID \citep{2022GCN.32657....1P}, photons of $\approx10$ GeV seen more than two weeks after the GRB by DAMPE \citep{2022GCN.32973....1D}, the spectacular detection by LHAASO of thousands of VHE photons up to 18 TeV \citep{2022GCN.32677....1H}, and potentially even a 250 TeV photon detected by Carpet-2 \citep{2022ATel15669....1D}.
 
 The burst caused a Sudden Ionospheric Disturbance \citep{2022GCN.32744....1S,2022GCN.32745....1G,2022RNAAS...6..222H,2023Atmos..14..217P}. There were no detected neutrinos associated with GRB 221009A, however \citep{2022GCN.32665....1I,2022GCN.32741....1K,2022arXiv221014116A}. The gravitational-wave detectors were off or not sensitive enough to achieve any detection \citep{2022GCN.32877....1P}.

GRANDMA (Global Rapid Advanced Network for Multi-messenger Addicts) \citep{GRANDMAO3A,GRANDMAO3B,2022MNRAS.515.6007A} is a collaboration of ground-based facilities dedicated to time-domain astronomy, and focused on electromagnetic follow-up of gravitational-wave candidates and other transients such as GRBs.
Its network contains 36 telescopes from 30 observatories, 42 institutions, and groups from 18 countries\footnote{\url{https://grandma.ijclab.in2p3.fr}}.
The network has access to wide field-of-view telescopes ([FoV] $>1\textrm{deg}^2$) located on three continents, and remote and robotic telescopes with narrower fields-of-view.

Here we present the analysis of the afterglow emission of GRB 221009A with different model approaches. All results are obtained using the \textit{Fermi} GBM trigger time of 9 October 13:16:59.99 UT.
In \S\ref{data}, we present the observational data we use in the article, the photometric methods we use and a discussion of the extinction selection. In \S\ref{dataanalysis}, we present our methods to analyze the afterglow light curves using empirical light-curve fitting and two Bayesian inference analyses. We then present our results to investigate which astrophysical scenarios and processes best describe the data. In \S\ref{conclusion}, we present our conclusions. 

\section{Observational data}
\label{data}
\subsection{\textit{Swift} XRT and \textit{HXMT}/LE afterglow data}\label{sec:swift_afterglow}
\begin{figure}[!th]
\centering
\includegraphics[width=1.0\columnwidth]{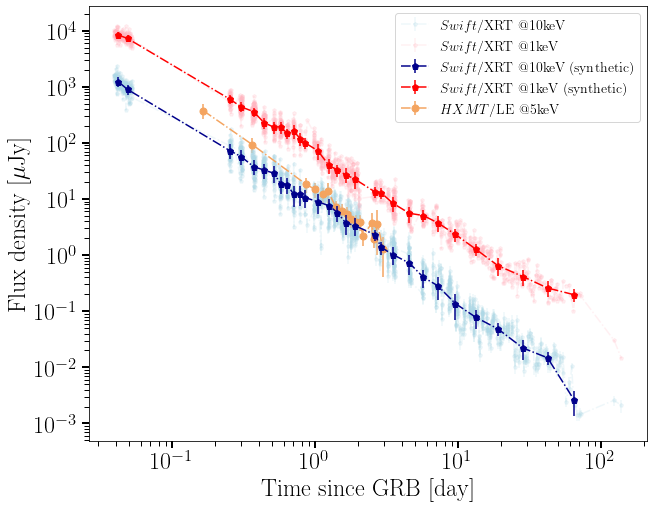}
  \caption{The unabsorbed X-ray light curve of GRB 221009A detected by the \textit{Swift}/XRT (given at 10 keV in blue and 1 keV in red) and the \textit{HXMT}/LE (orange) instruments. The light curves were corrected for Galactic and intrinsic N\ion{H}{1} column density absorption estimated from the late-time \textit{Swift}/XRT spectrum analysis (\url{https://www.swift.ac.uk/xrt_spectra/01126853/}). In dark blue and red colors, we show our synthetic \textit{Swift}/XRT light curve that we finally used in our afterglow modeling analysis, see the sections \ref{IAP} and \ref{NMMA}.}
   \label{fig:x-ray_LC}
\end{figure}

%\subsubsection{\textit{Swift}/XRT}
The \textit{Swift} XRT started to observe the field of GRB 221009A right after BAT triggered on the afterglow, about 56 min after the \textit{Fermi}/GBM trigger time. The X-ray light curve ($0.3-10$ keV) of GRB 221009A was collected from the UK \textit{Swift} Science Data Centre\footnote{\url{https://www.swift.ac.uk/}} at the University of Leicester \citep{2007A&A...469..379E,2009MNRAS.397.1177E}. We directly made use of the Burst Analyser light curve given in Jansky units at the 10 keV central frequencies \citep{2010A&A...519A.102E}.
Due to the large number of data points in the \textit{Swift} XRT light curve, we could not use it directly for the MCMC analysis without overweighting the X-ray data. We, therefore, constructed a synthetic light curve of the \textit{Swift} XRT data (both at 1~keV and 10~keV). Assuming a power law spectrum within the \textit{Swift} XRT passband (as found by the \textit{Swift} spectral analysis\footnote{\url{https://www.swift.ac.uk/xrt_spectra/01126853/}}), the 1~keV band is constructed using an extrapolation of the 10~keV flux density curve at the times where the photon index could be derived from the \textit{Swift} burst analyzer hardness ratio analysis. We separated the observations in both bands into 29 time windows, fitting a Gaussian to the flux distribution of the observations in each time window. Its median value and standard deviation are used as the measure and error of the synthetic curve. The obtained synthetic light curve is presented in Fig. \ref{fig:x-ray_LC}.
\medbreak
%\subsubsection{HXMT/LE}

The \textit{Insight}-HXMT/LE X-ray telescope \citep{2020SCPMA..6349502Z}  detected the afterglow emission of GRB 221009A at late times from about 9.8 h to 3 d after the \textit{Fermi}/GBM trigger time, including two scanning observations (P050124003601 \& P050124003701) and 20 pointing observations ranging from P051435500101 to P051435500401 with a total good-time-interval of 24 ks. The first two points are obtained by the spectral fitting of two scanning observations. The spectrum is obtained from the data when the target appears in the FoV. Unlike the pointing observations, the background is not obtained by the background model but from a region with no bright source in the FoV. Moreover, the instrumental response is calculated with the target track in the FoV and the Point Spread Function (PSF) of the \textit{Insight}-HXMT/LE collimator. For the pointing observations, we use the \textit{Insight}-HXMT Data Analysis software HXMTDAS v2.05\footnote{\url{http://hxmtweb.ihep.ac.cn/}} to extract the light curves, spectra and background following the recommended procedure of the \textit{Insight}-HXMT Data Reduction for HXMT-LE analysis.  For both the scanning and pointing observations, the spectra of \textit{Insight}-HXMT/LE in the $1.5-10$ keV range are fitted by an absorbed power-law, i.e., tbabs*power in XSPEC. The \textit{HXMT}/LE X-ray afterglow is shown in comparison to the \textit{Swift}/XRT measurements in Fig. \ref{fig:x-ray_LC}.

\medbreak
The \textit{HXMT}/LE flux measurements are not corrected for the dust echo scattering \citep[see for example,][]{2023arXiv230101798N,2023Swift,2023arXiv230211518T} since Insight-HXMT is a collimated telescope with a large field of view, e.g., $1^{\circ}\times 4^{\circ}$. It is thus not possible to easily remove  the dust echo scattering flux contribution. However, to evaluate the apparent flux increase caused by the dust scattering echoes we employed a similar procedure than the one adopted on the NICER data \citep{2023Swift}. As the dust scattering only dominates at energies below 4 keV, we first restricted the energy range to 4-8 keV. As a mild change of the spectral shape is observed during our \textit{Insight}-HXMT observations  \citep{2023Swift}, we used the spectral model of \textit{Swift}/XRT with a fixed Galactic/intrinsic $N_{\rm H}$ and photon index ($\Gamma$=1.8) to fit the spectra. For the first several pointing data, the derived unabsorbed fluxes above in the 1.5-10 keV energy range are then consistent with the interpolated fluxes of \textit{Swift}/XRT at the same times, which are about 10\% lower than the uncorrected/dust-echo-included fluxes. However, for the last several pointing data, the combination of the low counts rate of the source and the high background level prevents us from correcting the fluxes caused by dust scattering echoes. To keep all \textit{HXMT} fluxes produced in the same way, only the uncorrected dust echos scattering fluxes are given for all the data of LE. Therefore, these fluxes are systematically brighter (about 10\%) than that derived from \textit{Swift}/XRT at early (<0.5 d) times. We also notice that the flux difference between Insight-\textit{HXMT}/LE and \textit{Swift}/XRT narrows as time goes, which could be due to the fading of the dust scattering echoes.

%We expect that the effect of dust scattering echoes on the fluence measured by \textit{HXMT}/LE is less than in \textit{Swift} measurements \citep{2023Swift}, leading to different overall brightnesses in the light curve. This is because the sensitivity of \textit{HXMT}/LE is highest above a few keV, whereas for \textit{Swift}, the contribution from X-ray rings on the measured fluence is expected to be minimal.}

\subsection{Optical observations during the GRB prompt emission}
\label{sec:mundrabilla}
We used the images taken from two sites managed by the Desert Fireball Network \citep{2020PASA...37....8T} at 
Mundrabilla (lon = 127.8486$^{\circ}$ E, lat = 31.8356$^{\circ}$ S, altitude = 84\,m) and at Raw War Road (lon = 125.7503$^{\circ}$ E, lat = 29.7422$^{\circ}$ S, altitude = 215\,m), Western Australia. The acquisition devices are constituted of a Nikon\,D810 (color CFA matrix) set at 3200\,ISO with Samyang\, 8mm\,F/3.5 optics. This provides images covering the full sky. 
Images have 27\,s exposure time taken every 30 seconds for the entire night. At the prompt time the GRB is located at elevations of $15^{\circ}$ and $17^{\circ}$ above the local horizons of Mundrabilla and Raw War Road, respectively. The sky at Mundrabilla was partially covered by thin clouds and there was bright moonlight. Weather and elevation conditions were better at Raw War Road.
We analyzed the archive images taken between t$_{GRB}-30$ s and t$_{GRB}+500$ s. There is no detection 
at the position of the GRB to a limiting magnitude of 3.8 mag in the Green filter (which is roughly compatible with Johnson $V$) at Raw War Road. The limits are shallower at Mundrabilla. 
Time and magnitudes in the AB system are reported in Table~\ref{tab:GRANDMA_observations} corrected for extinction and in the Appendix, Table~\ref{tab:all_observations} uncorrected for extinction. No other contemporaneous observations have been reported, so to our knowledge, these are unique.

\subsection{Optical post-GRB observations}
\label{sec:post-grbobs}
Our first observation of the GRB within GRANDMA was obtained with the TAROT-R\'eunion telescope (TRE) at 2022-10-09T15:34:41 UTC (2:20\,hr after T$_0$) thanks to its automated program following GRBs. Although GRANDMA was not conducting an observational campaign at the time of event, by the request of A. de Ugarte Postigo, the GRANDMA network was activated to observe about 1 day post-trigger time; at this point, we provided the network the \textit{Swift} UVOT coordinates \citep{2022GCN.32632....1D}. The first ToO image requested by GRANDMA was taken by the 60 cm telescope from Maidanak $\sim90$ min after the notification at 2022-10-10T14-56-43 UTC (1.08 d after the GBM trigger) with the $R_C$ filter. Our last observations were made by the Canada-France-Hawaii Telescope (CFHT) equipped with Megacam at 2022-10-29T06:32 (19 d, 17 hr post T$_0$). In total, we collected about 80~images (usually consisting of stacks of short exposures) from 15 GRANDMA partner telescopes. In successive order, we provide here the mid-time of the first observation relative to T$_0$ for each telescope and the filters used during the whole campaign: D810 (before and during the prompt emission in $V$ band) at Mundrabilla and Raw War Road observatories, TAROT-R\'eunion (0.0972 d, without filter) near Les Makes Observatory, UBAI-ST60 (1.0813 d in $R_C$) at Maidanak Observatory, KAO (1.1368 d in $g^\prime r^\prime i^\prime z^\prime$) at Kottamia Observatory, ShAO-T60 (1.1465 d in $VR_C$) at Shamakhy Observatory, AZT-8 (1.2274 d in $R_CI_C$) at Lisnyky observatory, HAO (1.2563 d without filter) at Oukaimenden observatory, MOSS (1.2722 d without filter) at Oukaimenden Observatory, C2PU-Omicron (1.3077 d in $r^\prime$) at Calern observatory, SNOVA (2.1535 d without filter) at Nanshan Observatory, T70 (2.2424 d in $I_C$) at Abastumani Observatory, UBAI-AZT22 (11.1313 d in $R_C$) at Maidanak Observatory, VIRT (12.4567 d in $R_CI_C$) at Etelman Observatory, and CFHT-Megacam (19.6945 d in $g^\prime r^\prime i^\prime z^\prime$) at Mauna Kea Observatory.
Our preliminary analysis of the GRANDMA observations has been reported by \cite{2022GCN.32795....1R} where we reported observations from UBAI-ST60, KAO, Lisnyky-AZT-8, MOSS, C2PU-Omicron and SNOVA. In general, the sensitivity of the observations at the earliest epochs was reduced by the full moon.
%\medskip

In addition to the professional network, GRANDMA activated its \href{http://kilonovacatcher.in2p3.fr/}{Kilonova-Catcher} (KNC) citizen science program for further observations. Our web portal was used to provide coordinates of the \textit{Swift} UVOT source. Some amateur astronomers participating in the program observed the source by their own volition and distributed their own reports to the astronomical community \citep{2022GCN.32640....1B,2022arXiv221212543R,2022GCN.32664....1R,2022GCN.32679....1R,2022GCN.32934....1A}. They also transferred their images to our web portal to allow us to perform our own image reduction and analysis. In total more than 250 images were uploaded to our web portal. Here we provide the list of the names of the telescopes (see Tables \ref{tab:GRANDMA_observations} and \ref{tab:all_observations} for the images selected for photometric analysis) : a Celestron C11-Edge telescope, iT11 and iT21 iTelescopes, the IRIS 0.68m telescope, the Celestron EdgeHD14, the 12" MEADE telescope at the RIT Observatory, the C11 Dauban MSXD Telescope, the 0.53-m Ritchey-Chretien telescope of Montarrenti Observatory, the OMEGON200F5Newton telescope, a Newton SW 200/1000 telescope, a Newton 250 f/4 telescope, a Celestron 11 ATLAS telescope, the T-CAT telescope at the Crous des Gats Observatory, the iT24 iTelelescope of the Sierra Remote Observatory and the 0.61-m Dall-Kirkham telescope of Burke-Gaffney Observatory, the 0.28m Mol SCT, the LCO 0.4m telescope at the McDonald Observatory, a Celestron C11 Millery telescope; the Planewave CDK-14 telescope at the Contern Observatory. The observations started 0.25 to $\sim6$ days after the trigger time, predominantly in Johnson-Cousins and Sloan filter sets, but also with other filters, such as $Lumen$ or Bayer sensors.

The GRANDMA observations are listed in Tables~\ref{tab:GRANDMA_observations} and \ref{tab:all_observations}. The former reports the mid-time (in ISO format with post-trigger delay) and extinction-corrected brightness (in AB magnitudes) of the observations, while the latter includes the uncorrected magnitudes and references to selected online GCN reports (see \href{https://gcn.gsfc.nasa.gov/other/221009A.gcn3}{public observational reports}, individual GCNs are all cited in the table). The mid-time is calculated as the weighted average of the observation start time and the number of exposures. The number of exposures is also provided. Our method for calculating magnitudes is described in the following section, and images that did not meet our criteria are labeled as ``VETO''. In Table~\ref{tab:all_observations}, the reference catalogs and stars used by external teams for comparison are also included, unless not specified in the GCN reports.  When the information is not provided by the online GCN report, we mark it as ``-''. 

\begin{figure*}%[!th]
\centering
\includegraphics[width=\textwidth]{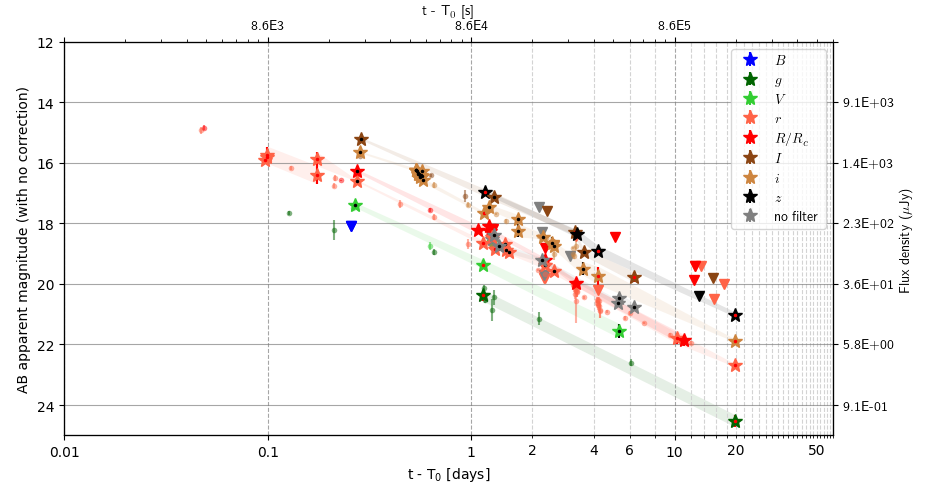}
  \caption{The optical afterglow of GRB 221009A was observed using $g^\prime Vr^\prime R_Ci^\prime I_Cz^\prime$ filters and without filter, with data points shown in the observer frame. The selected optical GCN data we use are represented by dots and the GRANDMA data measurements and upper limits are indicated by larger stars and downward-pointing triangles (see Table~\ref{tab:all_observations}). The red points within the stars indicate measurements made by professional observers, while black points represent observations made by KNC observers. Only magnitude measurements with at least a $3\sigma$ detection significance are included (the upper limits being given at $5\sigma$ significance), with uncertainty regions shown as shading. The measurements are not corrected for any extinction.}
   \label{fig:GRANDMAstars}
\end{figure*}

\subsection{Photometric methods}

We required all GRANDMA images to be pre-processed by the telescope teams with bias or dark subtraction and flat-fielding. We reject a few images from amateur astronomers where these corrections were not performed.  
Some teams uploaded their images with their own astrometric calibration, but for most images the astrometric calibration is obtained directly from the Astrometry.net website. Then, two methods are used to measure the magnitude on the template-subtracted images (see below):  
STDpipe and MUPHOTEN. For both of these methods, we use techniques to blindly search for new detections within the \textit{Swift} UVOT error localization \citep{2022GCN.32632....1D}, but we can also force photometry at the GRB 221009A afterglow coordinates we fixed to RA = 288.2646558, Dec. = 19.7733650 \citep{2022GCN.32907....1A}.

\begin{figure}[!th]
\centering
    \includegraphics[width=0.8\columnwidth]{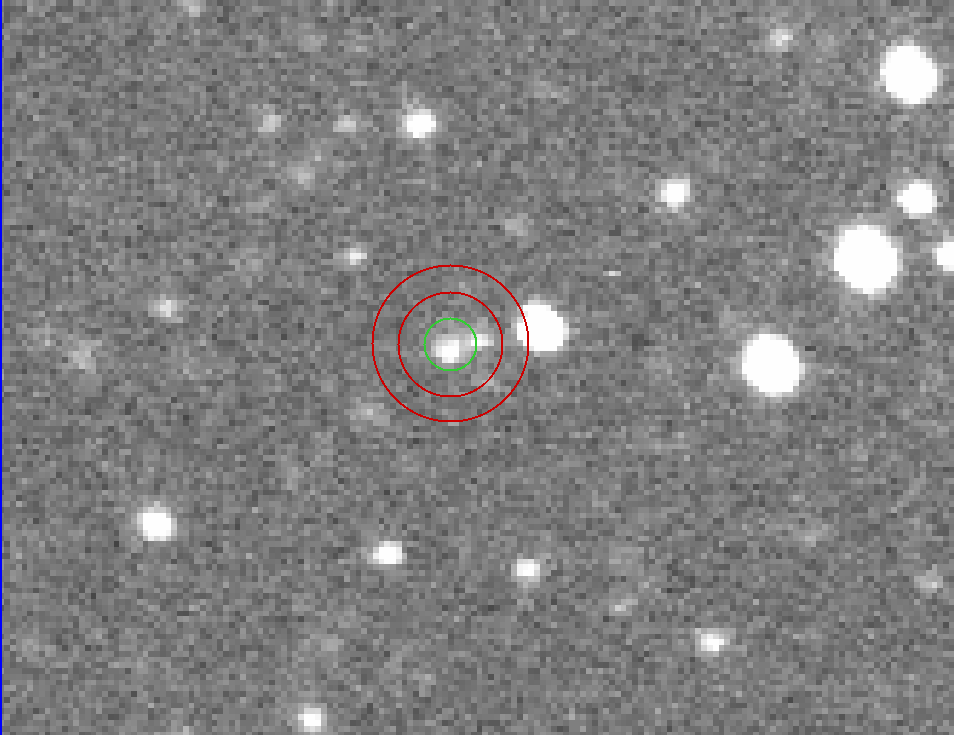}
    \includegraphics[width=0.8\columnwidth]{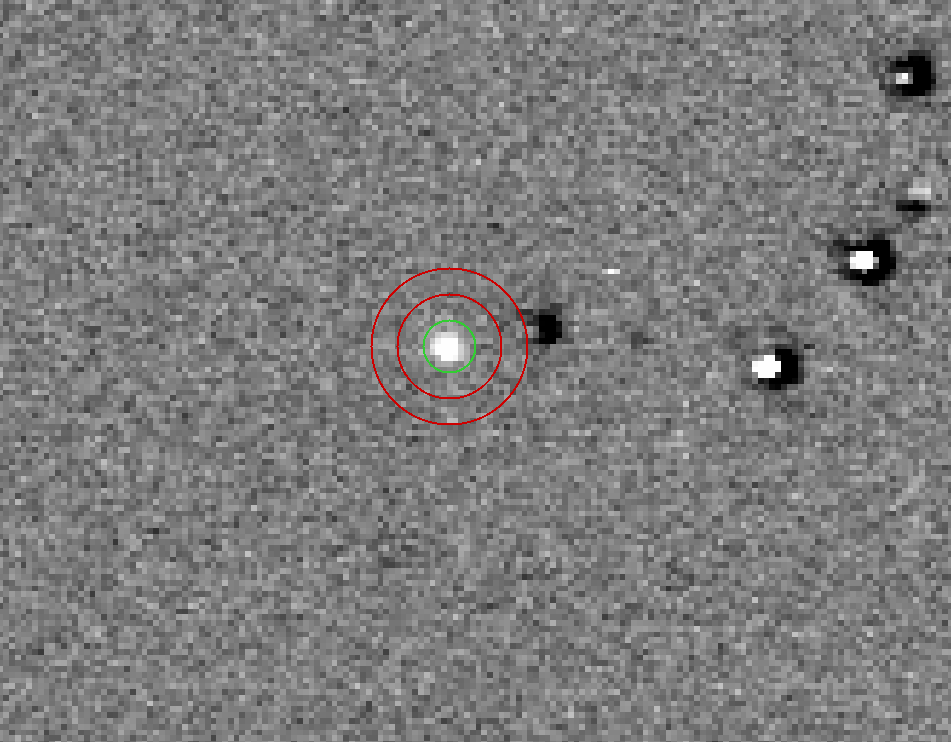}
    \caption{\textbf{Top panel:} A 60\,s exposure $I_c$ of GRB 221009A taken on the 0.61-m Dall-Kirkham458 telescope of the Burke-Gaffney Observatory $\sim$\,0.5 days after the Fermi GBM trigger time. \textbf{Bottom panel:} The corresponding difference image using a PS1 template image downloaded from the CDS {\sc HiPS2FITS} service. The optical afterglow is well-detected in the residual image and cleaned of any nearby star contribution. The aperture radii (the flux aperture in green and the background flux annulus in red) used to compute the photometric measurement are also shown in green in both images.}
    \label{fig:image_analysis}
\end{figure}

\textbf{STDpipe} --
The Simple Transient Detection Pipeline {\sc STDPipe} \citep{stdpipe}, is a set of python libraries aimed at performing astrometry, photometry and transient detection tasks on optical images. To do so, it uses several external algorithms such as {\sc SExtractor} \citep{Bertin96} for the source extraction, catalog cross-matching tools using the CDS Xmatch service developed at the Strasbourg Astronomical Observatory \citep{2012ASPC..461..291B,Pineau20}, the {\sc hotpants} code \citep{hotpants} for image subtraction tasks and the {\sc photutils}\footnote{\url{https://github.com/astropy/photutils}} Astropy package \citep{photutils} to perform photometric calibration and measurements. More details about the {\sc STDPipe} software architecture can be found in the git documentation\footnote{\url{https://github.com/karpov-sv/stdpipe}}. %From the KNC images, \DT{XX} of them did not satisfy our data quality criteria in STDpipe for scientific analysis. 
In order to increase the signal-to-noise ratio of some KNC images where the GRB afterglow was barely visible, we resampled and coadded individual frames using the {\sc Swarp} software \citep{2010ascl.soft10068B}.
Our final set of science images were subtracted with Pan-STARRS DR1 catalog \citep[PS1,][]{Chambers16} images downloaded from the CDS {\sc HiPS2FITS} service \citep{hips2fits} and rescaled to each image pixel scale. Forced aperture photometry was then applied at the GRB afterglow position in the residual images in order to limit the flux contribution from the very nearby stars. Due to the heterogeneity of the KNC instruments, we had a wide pixel scale distribution in our images. Therefore, the aperture radius was fixed per image to the average FWHM of stars detected at S/N $>5$ by {\sc SExtractor} in the image field. Depending on the photometric system used by KNC astronomers, the photometric calibration was done with the stars in the image field either using the native photometric bands ($g^\prime r^\prime i^\prime z^\prime $) of the PS1 catalog or by converting them into the Johnson-Cousins $BVR_CI_C$ system using the transformation described by \cite{2022A&A...664A.109P}. The photometric model for the calibrated magnitudes,$\textrm{mag}_{\textrm{cal}}$, is defined following method described in \cite{stdpipe}:
\begin{equation}
\textrm{mag}_{\textrm{cal}} = -2.5 \textrm{log}_{10}(\textrm{ADU})+\textrm{ZP}(\textrm{x},\textrm{y})+ \textrm{C} \times (\textrm{B}-\textrm{V})
\label{eq:photometry_model}
\end{equation}
where ADU is the star flux measured by the detector, ZP is the spatially varying Zero Point function and C is a color correction term to take into account the color distribution of the PS1 calibration stars. The ZP distribution is estimated by performing an iterative weighted linear least square fit to match the the photometric model given in Equation \ref{eq:photometry_model} to the catalogued magnitudes. 3$\sigma$ outliers to our photometric model were iteratively rejected (sigma clipping). The statistical errors on the ZP distribution were then propagated to the magnitude errors. Additional systematic errors due to the magnitude system conversion are also taken into account and can affect our measured magnitude up to 0.1 mag at maximum depending on the source brightness. Our KNC photometric results are shown in Table \ref{tab:GRANDMA_observations} and \ref{tab:all_observations}. As an illustration we show in Figure \ref{fig:image_analysis} one of the analysed KNC images according to the method described above.

%\midskip

\textbf{MUPHOTEN} --
\label{Muphoten}
{\sc Muphoten}\footnote{\url{https://gitlab.in2p3.fr/icare/MUPHOTEN}} is a Python-based software dedicated to photometry of transients observed by heterogeneous instruments, developed for the analysis of GRANDMA images \citep{2022PASP..134k4504D}. 
Similarly to {\sc STDPipe}, it uses Python libraries like {\sc photutils} \citep{photutils} and external algorithms like {\sc SExtractor} \citep{Bertin96} and {\sc hotpants} \citep{hotpants}. The {\sc Muphoten} software was utilized for the analysis of all GRANDMA images and a portion of KNC images. We first construct a template image by mosaicking Pan-STARRS DR1 (PS1) archive images, matching the image FoV, and we use {\sc hotpants} to subtract the template from the image. We do this to limit the contamination from nearby objects, in a field-of-view crowded due to its proximity to the Galactic plane.  However, for a limited number of images, the template subtraction was unsuccessful due to non-convergence with {\sc hotpants}. Nevertheless, these images had adequate resolution to clearly distinguish the transient from neighboring sources, so they were retained for further analysis. The background is estimated using the method of {\sc SExtractor}, in a mesh of $150\times150$ pixels by default (smaller grids were applied for images with rapidly varying backgrounds). The background and its standard deviation are interpolated to each pixel of the image and subtracted to obtain the final result. Sources are detected by identifying clusters of at least five neighboring pixels that exceed a threshold of $2\sigma$ above the background. %CFHT riz, KAO 10/10, MOSS, UBAIAZR22, VIRT

Next, we conducted isophotal photometry on all detected sources, measuring the flux and its corresponding error (obtained by integrating the squared flux error, computed by {\sc photutils} as background variation plus gain-adjusted Poisson noise, over the same elliptical aperture). The sources were cross-matched with the PS1 catalog, yielding the PS1 magnitudes of the matched sources in the corresponding filter. For images taken with Johnson-Cousins filters, we transformed the PS1 magnitudes to the observed filters using the conversion equations from \cite{2018BlgAJ..28....3K}. Unfiltered images were treated as if they were taken with the Cousins $R_C$ filter and were processed using the same conversion equations. We construct a calibration scale by fitting the instrumental magnitude and PS1 magnitude using a first-order polynomial fit with iterative clipping of outliers ($3\sigma$ away from the fit). We then compute calibrated magnitudes for all detected sources, sort them by distance to expected transient coordinates, and consider a source a detection if its coordinates match within five pixels. We also compute, for each source, the photometric error, adding contributions from the flux error measured above and from the calibration uncertainties.
Due to crowding in the Galactic plane, we checked for neighboring objects affecting the automatically computed apertures, reducing them if necessary.
Forced photometry using circular apertures of default radius 1.5 times the average FWHM of stars in the image was performed at the GRB coordinates in the absence of direct detection. This was calculated using the {\sc PSFex} software \citep{2011ASPC..442..435B}.
Plotting circular apertures of increasing radius (1 to 10 pixels) and their corresponding measured fluxes, we could check whether the default aperture collected all the transient flux and not neighbouring sources, and manually correct its coordinates and radius when needed. %transient coordinates and 1.5*psf radius by default

Finally, {\sc Muphoten} assesses the sensitivity of the image with upper-limit estimations. In {\sc Muphoten}, upper limits are computed as global properties of the whole studied image. The default method outlined in \citet{2022PASP..134k4504D} calculates the success rate of recovering PS1 objects based on 0.2 magnitude intervals and selects the faintest interval where more than 10\% of PS1 objects in the field-of-view are detected in the image. In the case of images where there is a high detection rate up until the limit of the Pan-STARRS catalog, an alternative method defines the upper limit as the magnitude of the faintest source detected with a SNR~> 5.

\subsection{Extinction selection}

Unfortunately for optical studies, the brightest GRB ever detected lies behind significant extinction near the Galactic plane ($b=4.32^\circ$). Following the maps of SF11, the line-of-sight at the ``reference pixel'' lies behind $E_{(B-V)}=1.32$ mag/$A_V=4.1$ mag. However, at Galactic latitudes $|b|<5^\circ$, the maps of \citet[][which those of SF11 are based on]{Schlegel1998} are known to be unreliable and may overestimate the extinction \citep[][and references therein]{2003AJ....126.2910P}.

\citet[][henceforth RF09]{2009MNRAS.395.1640R} presented a method of determining extinction towards the Galactic plane using near-infrared color excess determinations based on 2MASS observations, following earlier work from \cite{2005A&A...432L..67F} based on stellar counts. Using their extinction calculator\footnote{\url{https://astro.kent.ac.uk/~df/query_input.html}} and the position of the GRB, we find a significantly lower extinction, using the 100NN (Nearest Neighbour, see RF09 for details) result (which has the highest S/N), of $A_V=2.195$ mag. Using the classical Milky Way extinction curve of \citet[][henceforth CCM89]{1989ApJ...345..245C}, this translates into $E_{(B-V)}=0.709$ mag. The extinction maps of RF09 show that extinction toward this region of the Milky Way is smooth and quite homogeneous for several degrees around (and not high in the context of the potential extinction toward the Milky Way), the nearest pronounced molecular clouds with significantly higher extinction lie closer to the plane in the neighboring constellation Vulpecula, about $5^\circ$ away. Therefore, we deem the use of the extinction curve of CCM89 to be valid.

%Is this extinction value more realistic than that of SF11? 
The method of RF09 extends only\footnote{\cite{1980A&AS...42..251N} give a value of $A_V=3.3$ mag along this sightline out to 3 kpc.} to $2-3$ kpc. There is evidence for additional dust screens at larger distances, however. \textit{Swift} XRT observations reveal expanding rings in the X-rays \citep{2022GCN.32680....1T} arising from scattering on distant dust curtains. These authors report the discovery of nine dust rings and derive the distances, with the most distant one lying at $3635\pm36$ pc, potentially already beyond the detection range of the RF09 method. Observations with IXPE \citep{2023arXiv230101798N} confirm the most distant dust ring found by \textit{Swift} at $3.75\pm0.0375$ kpc, and report an even more distant dust curtain at $14.41\pm0.865$ kpc. Recently, \cite{Vasilopoulos2023} reported a detailed analysis of \textit{Swift} XRT data and also find evidence for dust out to 15 kpc (see \citealt{2023Swift} for further analysis), and \textit{XMM-Newton} analysis of a total of 20 dust rings presented by \cite{2023arXiv230211518T} detects an even further dust curtain at $\approx18.6$ kpc.

The Galactic disc exhibits a warp \citep[e.g.,][and references therein]{2014A&A...569A.125H}. The map derived by \citet[][their Fig. 16]{2014A&A...569A.125H} shows that at the Galactic longitude of GRB 221009A ($l=52.96^\circ$), HII regions indeed extend up to several hundred pc ``above'' the Galactic plane. For the Galactic latitude of GRB 221009A ($b=4.32^\circ$), the sightline lies $\approx1100$ pc above the plane at a distance of 14.4 kpc, beyond the HII regions mapped by \cite{2014A&A...569A.125H}. However, this does not rule out the existence of cold dust curtains even that high above the Galactic disc which would contribute extra extinction beyond the RF09 measurement. We therefore conclude the true extinction value along the line of sight to GRB 221009A lies in the interval of $A_V=2.2-4.1$ mag, and will discuss both extreme values.

\section{Multi-wavelength analysis of the afterglow}
\label{dataanalysis}

To analyze the afterglow light curve, we use data from multiple sources: Our own GRANDMA and KNC data (see Table \ref{tab:GRANDMA_observations}), selected GCN data (see Table \ref{tab:all_observations}), as well as data published in \cite{2023Swift,2023Shrestha,2023Laskar,2023arXiv230207761L,2023arXiv230207906O}. We especially note we use the \textit{Hubble Space Telescope} (\textit{HST}) data from \cite{2023arXiv230207761L} where the host-galaxy contribution has been subtracted using \texttt{galfit}. Near-infrared observations are taken from \cite{2022GCN.32654....1D,2022GCN.32755....1D,2022GCN.32758....1H,2022GCN.32804....1F,2023arXiv230207906O}, as well as the \textit{James Webb Space Telescope} (\textit{JWST}) MIR $F560W$ data point \citep{2023arXiv230207761L}.

\subsection{Empirical Light-Curve Analysis}
\label{empiricallc}

With the exception of our shallow upper limits from Mundrabilla and Raw War Road, no optical observations have been reported before the \textit{Swift} trigger.

The first observations, consisting of \textit{Swift} UVOT data from \cite{2023Swift,2023Laskar} and obtained via automatic analysis\footnote{\url{https://swift.gsfc.nasa.gov/uvot_tdrss/1126853/index.html}} as well as some ground-based observations \citep{2022GCN.32645....1B,2022GCN.32647....1X}, are found to decay more steeply than following observations (see Figure~\ref{fig:LC_ALL} in the Appendix), and also lie above the back-extrapolation of that data. This indicates an extra component in the light curve, potentially the tail end of a reverse-shock flash. The extreme intensity of the GRB makes it potentially possible that the early transient was extremely bright.

Fitting a joint multiband fit to the data, which assumes achromatic evolution and leaves only the normalization of each band an independent parameter, we derive a first decay slope of $\alpha_{steep}=1.32\pm0.34$ (we define $F_\nu\propto t^{-\alpha}\nu^{-\beta}$), significantly steeper than the later decay observed from $\sim0.09<t<0.59$ d, but quite shallow for a reverse-shock flash \citep{1999ApJ...520..641S,2000ApJ...545..807K}. As the baseline is short, it is possible we are seeing the transition from the early, steeply decaying component to the later shallower light-curve decay, and the decay at even earlier times might have been steeper and more in accordance with a reverse-shock flash. Extrapolating this slope backward to the peak of the brightest gamma-ray flare of the prompt emission, at $\approx220$ s post-trigger, we find $R_{AB}\approx11$ mag ($R_{AB}\approx7.6$ mag when corrected for SF11 extinction). This value is far fainter than our Mundrabilla/Raw War Road exposures probe. A steeper decay (see as mentioned early in the paragraph) or an additional component directly associated with the prompt emission cannot be ruled out but would still be unlikely to be bright enough to be detected by our shallow all-sky observations.

Data at $>0.09$ d can be fit with a smoothly broken power-law, with parameters pre-break slope $\alpha_1$, post-break slope $\alpha_2$, break time $t_b$ in days, and break smoothness $n$. The very last data points at $\gtrsim30$ d show a flattening that may result from the host galaxy becoming dominant, we exclude these data points from the analysis. We see no direct evidence of a SN component in the late light curve\footnote{Note that data presenting evidence of a photometric SN rise \citep{2022GCN.32769....1B,2022GCN.32818....1B} were taken under inclement conditions and are likely the result of blending with nearby sources and are therefore too bright (A. Pozanenko, priv. comm.). However, \cite{2023arXiv230111170F} assume an intrinsic optical decay slope identical to the X-ray slope and interpret the more shallow decay as a rising, luminous SN component.}, in agreement with \cite{2023Shrestha}, similar to the case of GRB 030329 \citep[e.g.,][]{2006ApJ...641..993K}, and therefore also do not include such a component in the fit. A dedicated search will need well-calibrated late-time data. In general, the data shows dispersion, leading to a large $\chi^2$.

This fit results in $\alpha_1=0.722\pm0.012$, $\alpha_2=1.437\pm0.003$, with $\sim$ $t_b=0.6$, and a sharp break $n=100$ fixed.%, with $\chi^2/d.o.f.=6.95$ \DT{same for the chisquare...}. 
Note that the errors of the fitted parameters are statistical only and do not include the systematic uncertainties ($\sim$10\%) due to the inter-calibration of the different photometric bands. Therefore, they are simply presented for diagnostic purposes. This steepening had also been reported by \cite{2022GCN.32755....1D}, who found $\alpha_1\approx0.8$, $\alpha_2\approx1.6$, and $t_b\approx0.98$ d based on a significantly smaller data set. \cite{2023Shrestha} find 
$\alpha_1=0.64$, $\alpha_2=1.44$ in $r^\prime$ and $\alpha_1=0.81$, $\alpha_2=1.46$ in $i^\prime$, similar to our result. \cite{2023Swift}, using only \textit{Swift} UVOT data, find $\alpha_{1,O}=0.98^{+0.05}_{-0.11}$, $\alpha_{2,O}=1.31^{+0.07}_{-0.05}$, and $t_{break,O}=0.255^{+0.197}_{-0.127}$ d, in agreement with our results within $2\sigma$. They point out this decay is clearly slower than that of the X-rays (see below), but is very unlikely to be influenced by a host or SN component.

\textit{Swift} XRT observations (initially reported in \citealt{2022GCN.32651....1K,2022GCN.32671....1T}, but these reports are based on the \textit{Swift} trigger time) as given in the XRT repository \citep{2007A&A...469..379E,2009MNRAS.397.1177E} show the light curve\footnote{\url{https://www.swift.ac.uk/xrt_live_cat/01126853/}} to have multiple shallow breaks (see also \citealt{2023Swift}, who caution that especially during the WT mode observation, the dust-scattering rings can influence the light curve stemming from the atypical background around the afterglow PSF), but within the first $\approx10$ d, the decay slope is $\alpha_X\approx1.5-1.6$, similar but steeper than our optical result. In their detailed analysis, \cite{2023Swift} find $\alpha_{1,X}=1.498\pm0.004$, $\alpha_{2,X}=1.672\pm0.008$, and $t_{break,X}=0.914^{+0.127}_{-0.116}$ d. \textit{Insight}-HXMT observations (\citealt{2022ATel15703....1G}, see also \citealt{2023AnHXMT}) also yielded a somewhat steeper slope $\alpha_X\approx1.66$. NICER observations also find $\alpha_X\approx1.6$ \citep{2022GCN.32694....1I}. The significantly more shallow decay phase in the optical ($\alpha_O\approx0.83$) as well as the earlier break at $t_b\approx0.6$ d are not seen in X-rays at all. The optical light curve also shows a much stronger break with $\Delta\alpha_O=0.617\pm0.013$ vs. $\Delta\alpha_X=0.174\pm0.009$.

\subsection{Analysis of the Spectral Energy Distribution}
\label{SED}

\begin{figure}%[!th]
\centering
\includegraphics[width=\columnwidth]{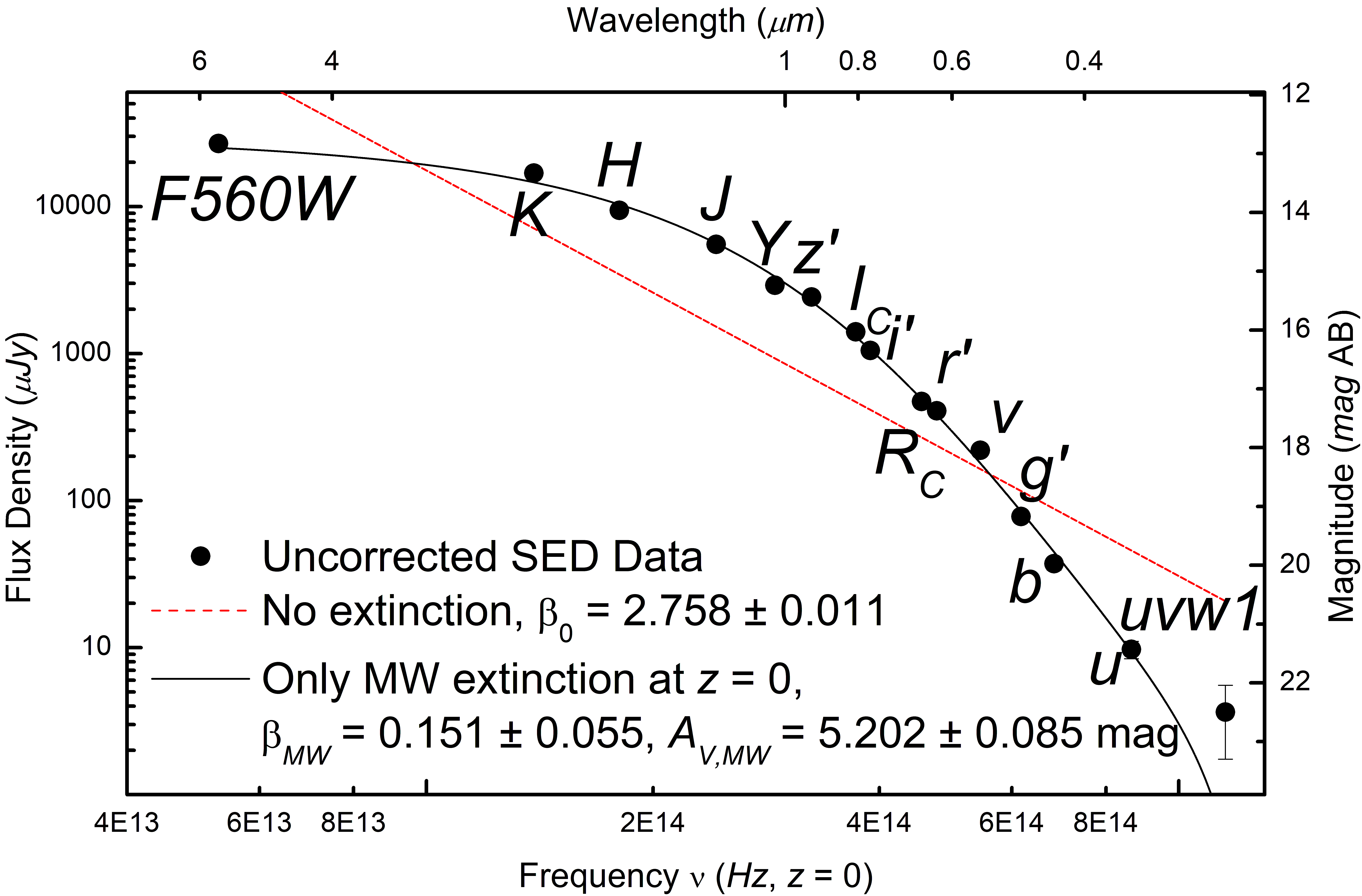}
\includegraphics[width=\columnwidth]{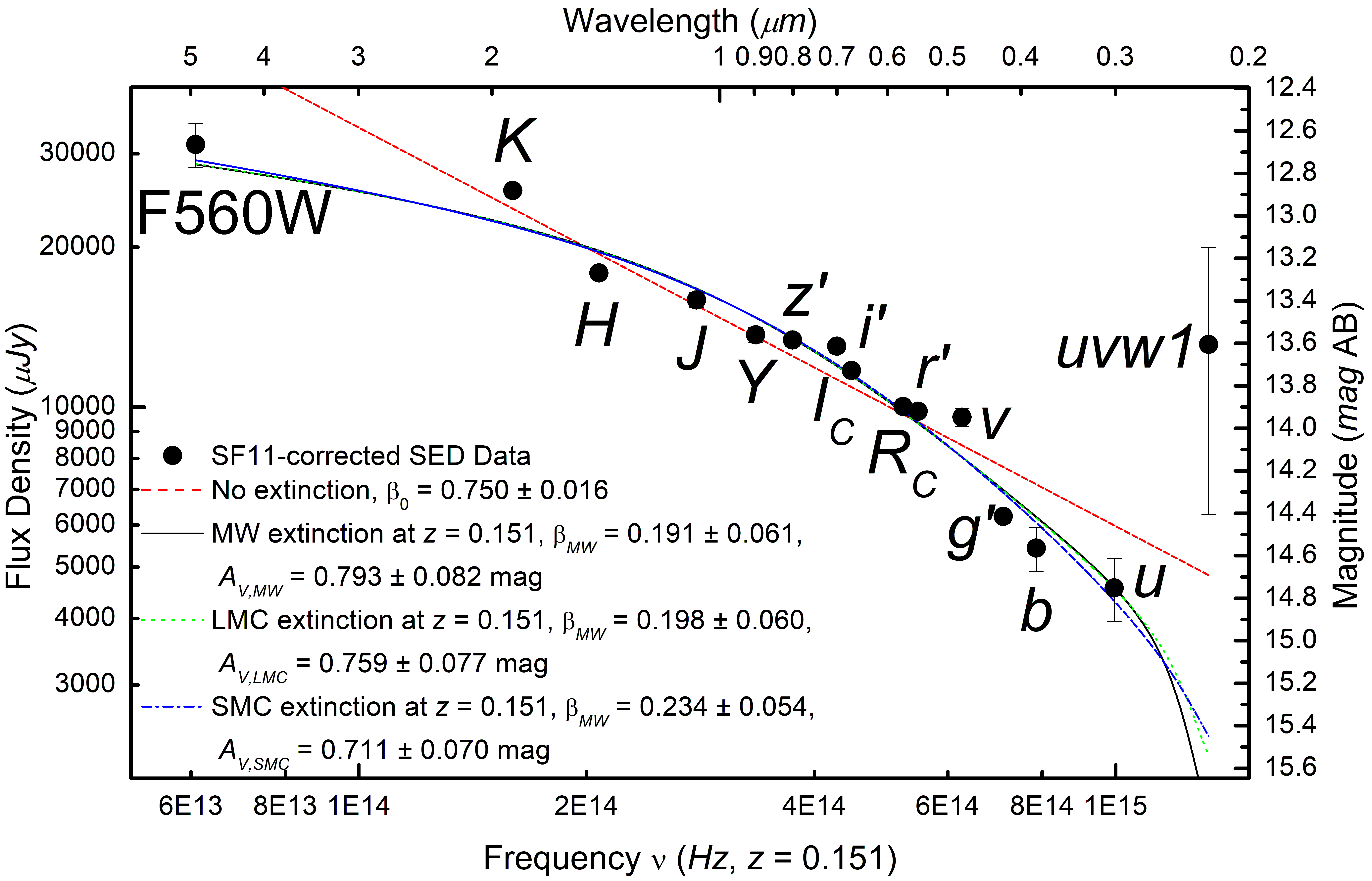}
\includegraphics[width=\columnwidth]{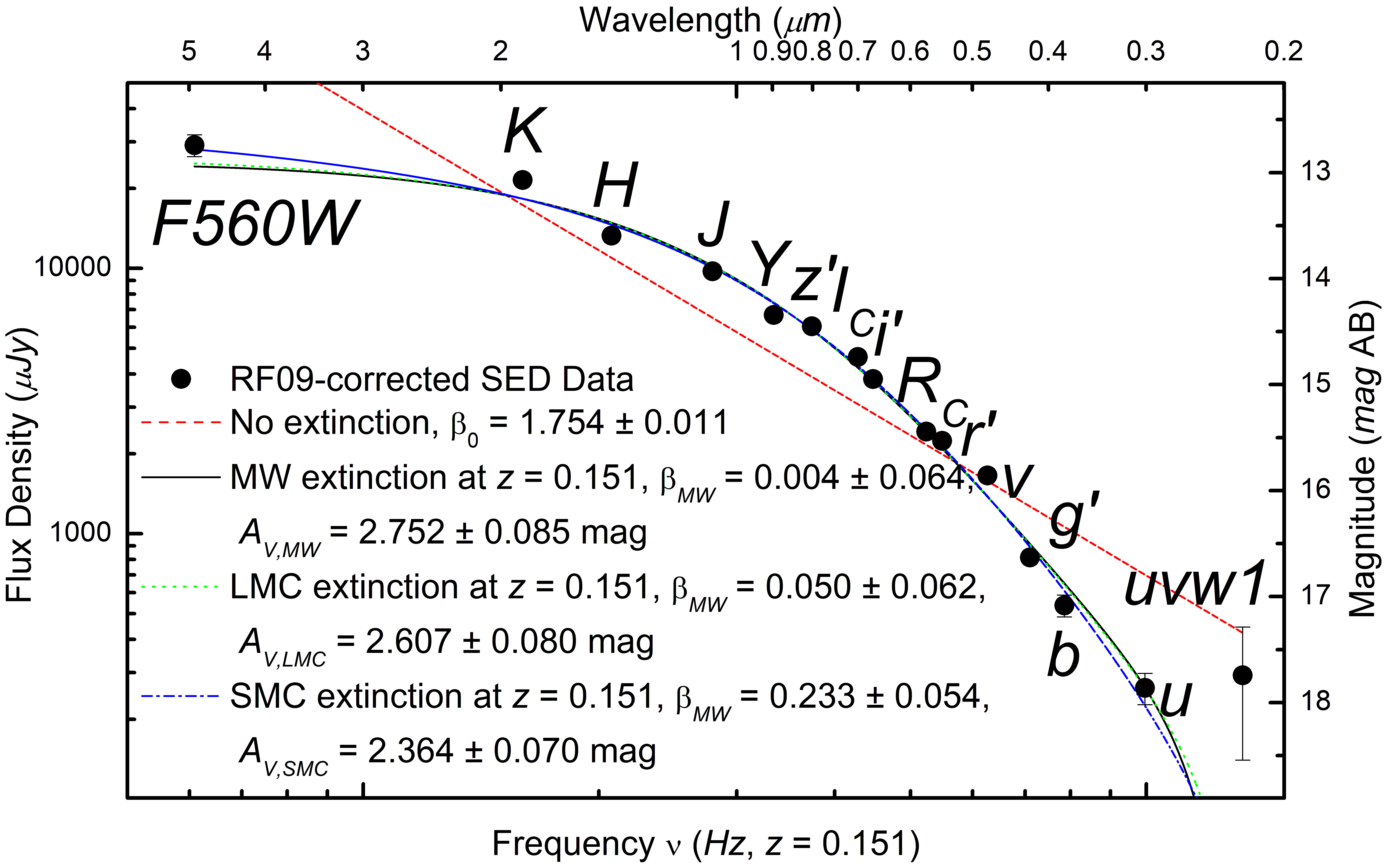}
  \caption{Analysis of the SED. \textbf{Top panel:} Fit to the uncorrected SED with a MW extinction model at $z=0$, i.e., assuming no additional host-galaxy extinction. The fit is generally in agreement with the data, with the $uvw1$ data point being brighter than the model. 
  \textbf{Middle panel:} Fit to the SED after correcting it for SF11 foreground extinction and shifting it to $z=0.151$. The correction leads to significant scatter, with the $uvw1$ now being clearly brighter than any fit, even one without additional host-galaxy extinction. The three extinction laws can not be discerned from each other, but the potential bright $uvw1$ emission makes the SMC law the preferred one. There is still significant curvature which implies additional significant host-galaxy extinction.
  \textbf{Bottom panel:} As the middle panel, with RF09 foreground extinction. The SED remains very red, and very high host extinction is implied. SMC extinction again leads to the most physical solution, but all dust laws are in conflict with the bluest \textit{Swift} UVOT detection.}
   \label{fig:SED}
\end{figure}

The normalizations derived from the joint multiband fit described in \S\ref{empiricallc} yield a Spectral Energy Distribution, a very low-resolution ``spectrum'' of the afterglow that is nonetheless valuable to study the dust properties along the line-of-sight. The fit assumes achromaticity, i.e., no spectral evolution, and is therefore based on all data involved in the fit. Except for scaling, the SED is identical at any time point covered by the fit, the specific values are measured at break time. 
While our data does not indicate an obvious spectral evolution, such a break is hard to constrain due to the challenge of building an empirical model that describes the data well.

We fit the SED both without extinction (a simple power-law) as well as with Milky Way (MW), Large (LMC) and Small Magellanic Cloud (SMC) dust following the parametrization by \cite{1992ApJ...395..130P}. These fits are performed after correction for Galactic extinction, and we study both the RF09 and SF11 models.
The derived SED shows scatter, with especially the $z^\prime$ band deviating and being too faint. The field is not covered by the Sloan Digital Sky Survey \citep[e.g.,][and references therein]{2023arXiv230107688A}; however, many telescopes use filters which are close to the SDSS system. There are offsets to the Pan-STARRS system which was used for calibration in most cases. Following the Pan-STARRS to SDSS conversion of \cite{2012ApJ...750...99T}, we find $g^\prime_{PS1}-g^\prime_{SDSS}=-0.26$ mag, $r^\prime_{PS1}-r^\prime_{SDSS}=0.02$ mag, $i^\prime_{PS1}-i^\prime_{SDSS}=0.03$ mag, and $z^\prime_{PS1}-z^\prime_{SDSS}=0.13$ mag, i.e., small changes for $r^\prime i^\prime$ but more significant changes to $g^\prime$ and $z^\prime$. As we are unable to examine each measurement individually for more precise color terms, we just apply these offsets to the four data points in the SED, which leads to a marked reduction in scatter and $\chi^2$. However, scatter still remains, with especially the $H$ band being fainter than the models and $K$ band being brighter. The precise light-curve fit leads to these normalizations to have small errors, causing large $\chi^2$ values that are not formally acceptable even for fits that generally model the SEDs well. The source of this scatter is less easy to understand than for the light-curve data points themselves, e.g., in the case of the $H$ and $K$ bands, most data points are from the final analyses presented in refereed papers \citep[e.g.,][]{2023arXiv230207906O}, and the few GCN points have larger errors and do not disagree with the fit curves. Furthermore, these data span a long time period, e.g., from 0.23 d to 25.4 d for $H$ and from 4.4 d to 25.4 d for $K$. This would imply all data from multiple sources are systematically offset in the same manner. As we have no immediate solution to this issue, we will continue to work with these results despite the fits being formally rejected, noting the scatter is approximately symmetric around the SED fit curves and not due to a clear discrepancy between model fit and data (as is the case for the fits without extinction which disregard the curvature of the SEDs).

\subsubsection{Pure MW extinction}

We first study the SED without applying any MW foreground correction, and taking the data at $z=0$. The SED is very steep and shows evidence for curvature (see Fig. \ref{fig:SED}, top panel). A simple power-law fit yields a spectral slope $\beta_0=2.758\pm0.011$. This is clearly not a good model, we find $\chi^2=8102$ for 13 degrees of freedom.

Applying MW dust to the SED yields a highly significant improvement ($\chi^2=142.8$ for twelve degrees of freedom, this number of degrees of freedom is identical for all extinction fits), and we derive $\beta=0.323\pm0.086$, $A_{V,Gal}=5.202\pm0.085$ mag ($E_{(B-V)}=1.69\pm0.03$ mag). This value exceeds the SF11 correction by over a magnitude and can indicate two things: Either even the SF11 result does not encompass the entirety of the MW foreground extinction, or there is additional, significant host-galaxy extinction along the line-of-sight, or it is a combination of both at once. The detection of the \ion{Na}{1} doublet at the redshift of the host galaxy \citep{2022GCN.32648....1D,2023arXiv230207891M} indicates there must be some amount of host-galaxy extinction. However, we note that the free fit already yields an intrinsic spectral slope lying in the typical range found for GRB afterglows, $\beta\sim0.2-1.2$ \citep{2010ApJ...720.1513K}.

\subsubsection{SF11 extinction}

We next correct the SED for SF11 MW extinction and now study the pure host extinction at $z=0.151$. After this correction, the spectral slope is obviously much flatter than before ($\beta_{0,SF11}=0.750\pm0.016$, $\chi^2=207$ for 13 degrees of freedom), however, the SED shows remaining significant curvature as well as scatter (see Fig. \ref{fig:SED}, middle panel), with especially the $uvw1$ band deviating. We caution this color is derived using only two $r^\prime/R_C$-band GCN points, which yields additional uncertainty, beyond the fact that it is only a $2\sigma$ excess above the background, and thus barely a confident detection. If real, the $uvw1$ band detection \citep{2023Swift} coincides with the 2175 {\AA} bump feature for LMC and MW dust in the host-galaxy rest frame, indicating the host-galaxy extinction law is most likely similar to SMC dust which lacks this feature almost completely. Mathematically, the different results can not be distinguished ($\chi^2=128,\;125,\;116$ for MW, LMC, and SMC dust, respectively; see also \citealt{2023Swift}), but SMC dust yields the overall most logical result, with an intrinsic spectral slope very close to the MW-only fit ($\beta_{SF11,SMC}=0.234\pm0.054$) and moderately high additional host-frame extinction ($A_{V,SF11,SMC}=0.711\pm0.070$ mag, $E_{(B-V)}=0.243\pm0.024$ mag). In terms of the intrinsic spectral slope, our results are in good agreement with those of \cite{2023arXiv230207761L}, who give several results in the slope range of $\beta=0.3-0.4$, but especially a broadband fit using \textit{JWST} and Gran Telescopio Canarias spectra as well as NOEMA sub-mm data which yields $\beta=0.362$. They attribute most extinction to the Milky Way and find only very small host-galaxy extinction. Modeling these data together with XRT, they find evidence for a spectral break between the optical and X-ray bands, but not for a spectral break within the optical/NIR regime itself. Using only \textit{Swift} UVOT data, \cite{2023Swift} derive higher values: Correcting for the higher foreground extinction given by \cite{Schlegel1998} and using an intrinsic slope of $\beta=0.7$, they find $E_{(B-V)}=0.51\pm0.03$ mag for SMC dust.

\subsubsection{RF09 extinction}

Finally, for the lowest assumed MW extinction, that of RF09, we find a combination of ``moderately high'' MW extinction and ``moderately high'' host-galaxy extinction. The SED after RF09 correction is still very steep (see Fig. \ref{fig:SED}, bottom panel, we find $\beta_{0,SF11}=1.754\pm0.011$, $\chi^2=2300$). Again the three dust models yield similar goodness-of-fit values ($\chi^2=161,\;152,\;123$ for MW, LMC, and SMC dust, respectively, but in this case, the very flat intrinsic spectral slopes $\beta\approx0.0-0.1$ additionally speak against MW and LMC dust being the correct solution. SMC dust results in $\beta_{RF09,SMC}=0.233\pm0.054$, $A_{V,RF09,SMC}=2.364\pm0.070$ mag. Even this result is not in agreement with the $uvw1$ detection, however.

Overall, while there is no strong evidence for one or another foreground extinction, the most logical solution is SF11 foreground extinction with additional moderately high SMC extinction in the host galaxy. High host-galaxy extinction such as in the RF09 case is also not supported by the relatively small equivalent width of the Na doublet at the host redshift \citep{2023arXiv230207891M}. In general, given the poor performance of the fits, it seems like that the extinction law is different from the three canonical functions used above, or the spectrum cannot be approximated as a power-law; however, for the sake of the analysis, for both foreground extinction scenarios, we corrected our data for Galactic extinction and host-galaxy extinction. During this step, we accounted for the systematic uncertainties of our extinction estimate in each band, adding them to the photometric errors of Table \ref{tab:all_observations}. These fully-corrected data sets are shown in Table \ref{tab:GRANDMA_observations}.

\subsection{The afterglow of GRB 221009A in a global context - luminous but not intrinsically extraordinary}
\label{KannPlots}

\begin{figure}%[!th]
\centering
\includegraphics[width=\columnwidth]{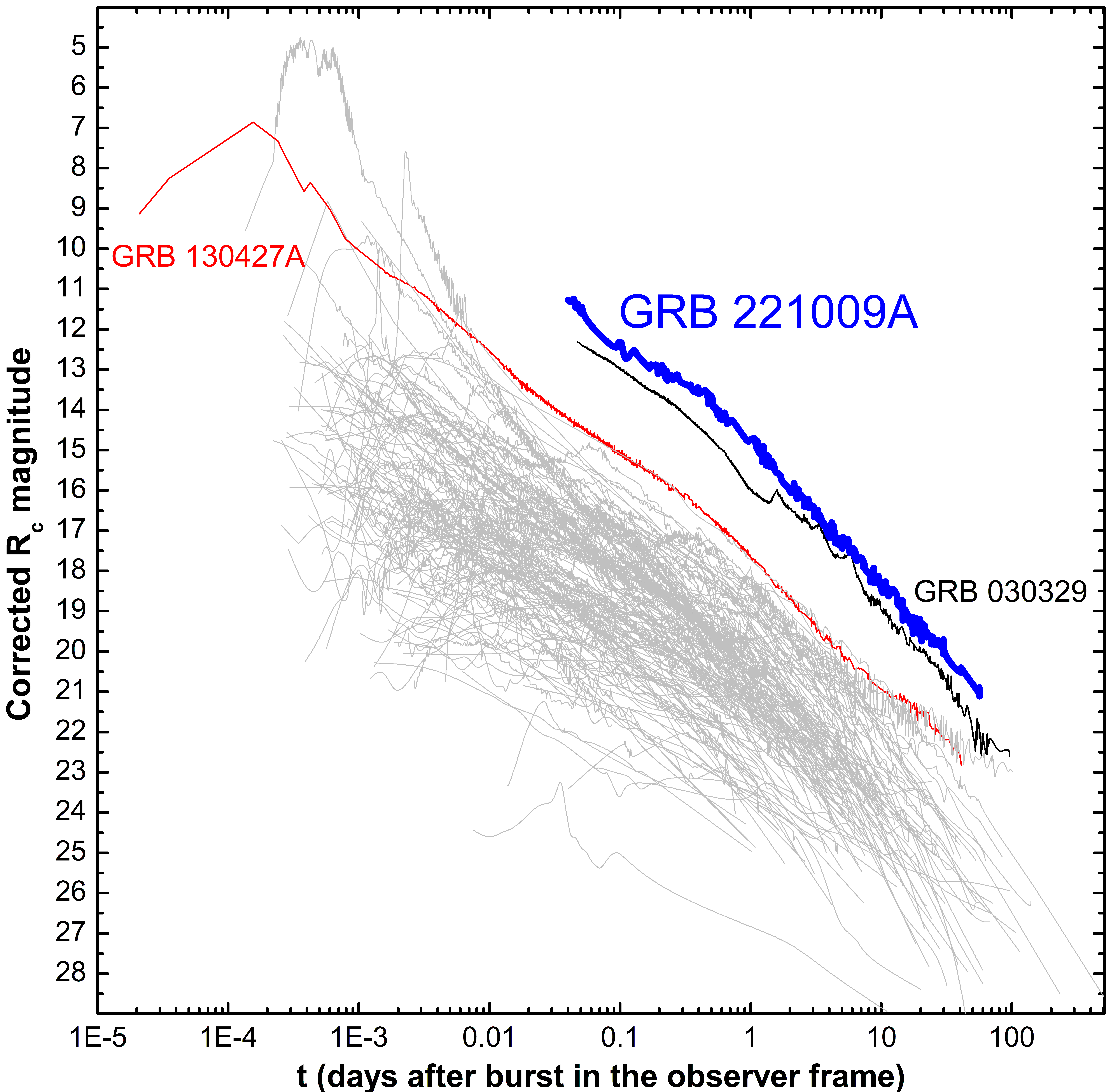}
  \caption{The afterglow light curve of GRB 221009A in context of a large sample of GRB afterglows \citep[][2023a,b in prep.]{2006ApJ...641..993K,2010ApJ...720.1513K,2011ApJ...734...96K}. These data have been corrected for Galactic extinction along each individual line-of-sight, and if possible for the host-galaxy and SN contribution. For the GRB 221009A afterglow, we show the result for SF11 Galactic extinction. We highlight the afterglows of two other GRBs, namely that of the much less energetic but similarly distant GRB 030329, and that of the well-studied, ultra-bright GRB 130427A, which had been the closest highly energetic (``cosmological'') GRB so far. Assuming the higher extinction correction, the afterglow of GRB 221009A is seen to be the brightest that has ever been detected, even brighter than the afterglow of GRB 030329, however, by only a small margin.}
   \label{fig:KP221009A_Obs}
\end{figure}

\begin{figure}%[!th]
\centering
\includegraphics[width=\columnwidth]{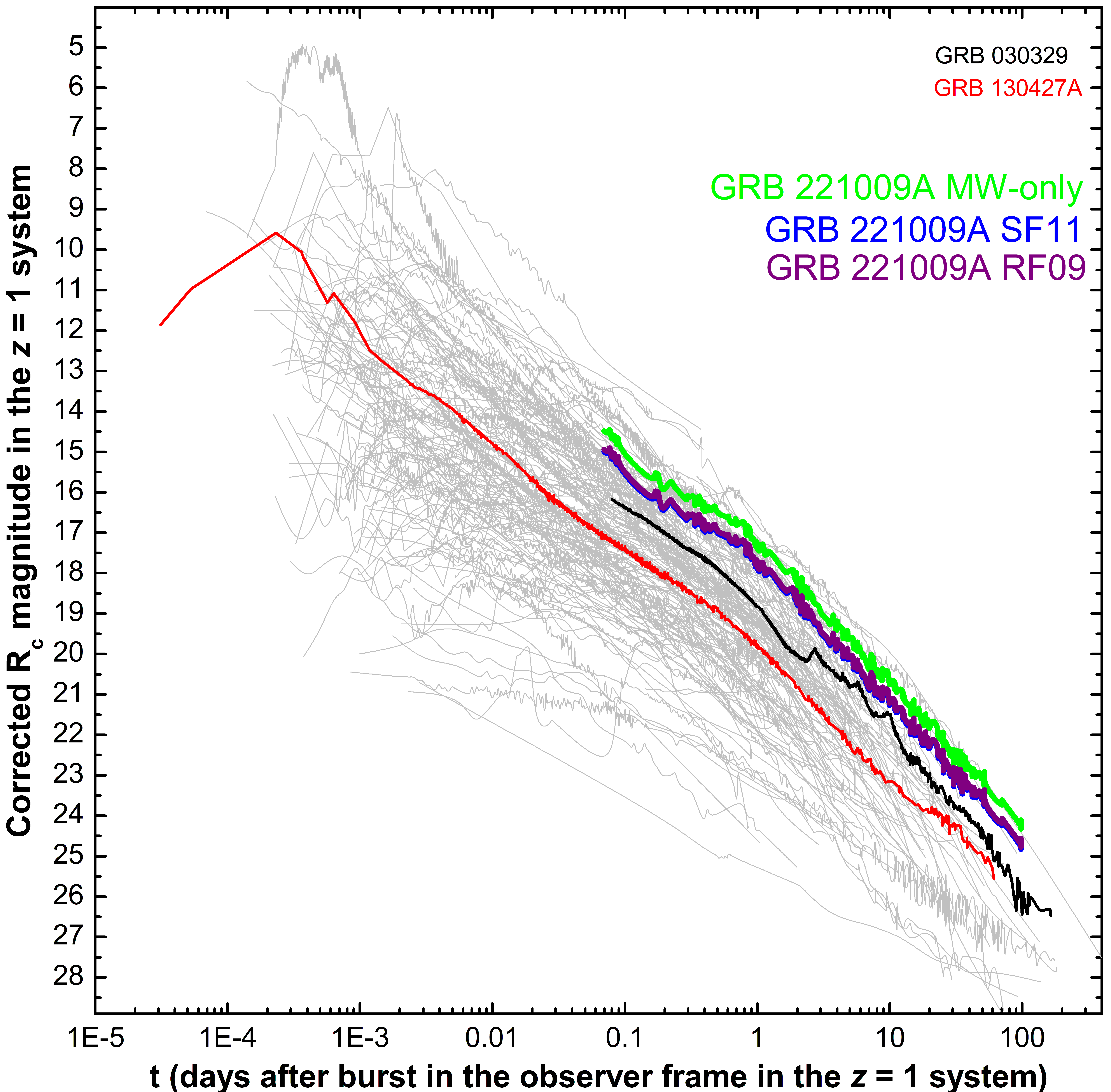}
  \caption{The same as Fig. \ref{fig:KP221009A_Obs}, but now all afterglows are in the $z=1$ system. This means that the afterglow magnitudes have been additionally corrected for host-galaxy extinction, and all of them have been shifted to $z=1$ taking the individual spectral slopes $\beta$ and cosmological $k$-correction into account. We again highlight the afterglows of the bright nearby GRBs 030329 and 130427A, and three solutions for GRB 221009A: The pure MW solution, SF11 MW extinction, and RF09 MW extinction along with the respective host-galaxy solutions. All yield similar brightness, with the SF11 and RF09 results essentially overlapping (offset by only 0.05 mag), and the afterglow is seen to be among the more luminous ones detected so far. We note that the late afterglow of GRB 221009A is \textit{not} corrected for a potential SN contribution and therefore the luminosity may be overestimated.}
   \label{fig:KP221009A_zone}
\end{figure}

With knowledge of the intrinsic extinction and the redshift, and using the method first presented in \cite{2006ApJ...641..993K}, we are able to place the optical/NIR afterglow of GRB 221009A in the context of a large sample of GRB afterglows. The sample is compiled from \citet[][Kann et al. 2023a,b in prep.]{2006ApJ...641..993K,2010ApJ...720.1513K,2011ApJ...734...96K}. These afterglows have been corrected for individual Galactic foreground extinction, host-galaxy contribution (where known) and SN contribution at late times (where applicable).

The otherwise as-observed light curves are shown in Fig. \ref{fig:KP221009A_Obs}. We highlight the afterglows of the two exceptional GRBs mentioned in the introduction. For one, the nearby but only moderately energetic GRB 030329, whose afterglow \cite[e.g.,][]{2004ApJ...606..381L,2006ApJ...641..993K} remains the most well-observed until this day, and is seen to be brighter than all other afterglows in the sample at any given time. And secondly the afterglow of the extremely bright GRB 130427A \citep[e.g.,][Kann et al. 2023b, in prep.]{2014Sci...343...38V,2014ApJ...781...37P}, also among the brightest observed GRB afterglows and energetically more similar to GRB 221009A.

The placement of the afterglow of GRB 221009A depends on the MW foreground extinction correction. From our three models, we display the SF11 solution here, which is usually the standard correction for extinction in other cases. If we use the MW-only solution, the resultant afterglow would be even brighter, whereas it would be magnitudes fainter with the RF09 solution, but as pointed out, this solution is unlikely. For SF11, we see the observed afterglow is even brighter than that of GRB 030329 at all times (albeit usually by not more than one magnitude) - potentially, yet another record that GRB 221009A holds. \cite{2023Swift} report the observed afterglow of GRB 221009A is by far the brightest X-ray afterglow, and also the brightest UVOT afterglow (after extinction correction) ever detected.

A better afterglow comparison can be achieved if we correct both for the distance (temporally and in terms of luminosity, we choose to place all afterglows at $z=1$ and present them in the observer frame) as well as for any intrinsic (host-galaxy) extinction. If the latter value is high, it can hide extremely luminous GRB afterglows from initially looking extraordinary \citep[e.g., GRB 080607,][]{2011AJ....141...36P}. The results are shown in Fig. \ref{fig:KP221009A_zone}. It can now be seen that the afterglow of GRB 130427A is only of medium brightness, and that of GRB 030329, while brighter, is also well within the sample of known afterglows. The same is true for the afterglow of GRB 221009A. The three foreground-extinction solutions yield similar results now, as high foreground extinction implies low additional host-galaxy extinction (MW-only, SF11), while the lower RF09 foreground extinction is mostly compensated by necessary high intrinsic extinction. Indeed, the degeneracy between the foreground and host-galaxy extinction, which stems from the very low redshift of the event, leads to the completely corrected light curves for the SF11 and RF09 extinction to almost overlap, the offset is only 0.05 mag. This motivates us to only use the SF11 solution for numerical modelling (see \S\S\ref{NMMA}, \ref{IAP}.)The afterglow of GRB 221009A is clearly among the more luminous detected so far, but it is not egregious. Only at late times does the unbroken decay lead it to become exceptional, but we caution these observations are not corrected for host-galaxy and SN contribution and are therefore to be taken with caution (see \citealt{2023arXiv230111170F} for the potential SN contribution, but see also \citealt{2023Shrestha}).

Quantitatively, we determine the $z=1$ magnitudes of a sample of 170 GRB afterglow light curves including that of GRB 221009A, where not every afterglow has measurements at the chosen time (however, the GRB 221009A afterglow does). As times, we chose 0.07 d, 0.5 d, 1 d, 5 d, 10 d, and 20 d. For these times, the comparison sample encompasses 140, 144, 130, 78, 52, and 34 other afterglows, respectively. We find the afterglow of GRB 221009A is brighter than 83.6\%, 86.1\%, 83.8\%, 87.2\%, 88.5\%, and 85.3\% of all other afterglows, respectively. More generally, it is brighter than $80\%-90\%$, which supports our claim that it is not exceptional in the way the prompt emission energetics are. However, it is extraordinary indeed in one aspect. As an example, at one day after trigger (12 hr in the rest-frame), there are 20 afterglows found to be brighter than that of GRB 221009A, but none of these lie at $z<1.4$, and 15 lie at $z>2$. In all time slices, the single afterglow at $z<1$ found to be brighter (at 0.5 d and 2 d, but not 1 d) is that of GRB 110715A at $z=0.8225$ \citep[][, Kann et al. 2023a, in prep.]{2017MNRAS.464.4624S}.

Overall, despite its extreme energetics, the optical/NIR afterglow of GRB 221009A is not intrinsically extraordinary compared to the global sample of known afterglows, a phenomenon also seen for other highly energetic GRBs like GRB 990123 \citep{2010ApJ...720.1513K}. \cite{2023Swift} reach similar conclusions, for both the UVOT and the X-ray afterglow, and \cite{2023Laskar} show this is true as well for the radio afterglow.

\subsection{Properties of the GRB afterglow from Bayesian Inference}\label{sec:modelling}

We analysed our data in the framework of the standard afterglow model where the observed emission is dominated by the synchrotron radiation from shock-accelerated electrons at the forward external shock due to the deceleration of a relativistic jet by the ambient medium (assumed here to be uniform). We explored the allowed parameter space of this model with a Bayesian approach using the constraints provided by two distinct data sets.
%We applied our numerical models to two multi-wavelength data sets. For both data sets, we use
Both data sets include X-ray
data from the \textit{Swift}/XRT instrument (at 1 keV and 10 keV), following the procedure outlined in \S\ref{sec:swift_afterglow} and downsampling the data to avoid our Bayesian inference runs being entirely dominated by X-ray observations. These data were combined with \textit{HXMT}-LE data
at $5\, \mathrm{keV}$.
%, which span $1.5-10$~keV (we take it to be ``5~keV''). 
We then combine the \textit{Swift}-XRT and \textit{HXMT} X-ray data with two different optical data sets:
(i) GRANDMA data points in the optical and near-infrared (presented in \S\ref{sec:mundrabilla} and \S\ref{sec:post-grbobs}), simply completed in u-band by early \textit{Swift}-UVOT points. We denote this set as \textbf{``GRANDMA''}.
%\textit{Swift}/UVOT and GRANDMA data presented in \S\ref{sec:mundrabilla}, \S\ref{sec:GRANDMAafterglow} and \S\ref{sec:mundrabilla}, $u$, $b$, $g^\prime$, $V$, $r^\prime$, $R_C$, $i^\prime$, $I_C$, and $z^\prime$ bands. We denote this set as \textbf{``GRANDMA''}.
(ii) We enrich the \textbf{``GRANDMA''} data with
the same observations collected in the literature
as already used in the introduction of \S\ref{dataanalysis}. This full data set has the advantage of including $J$, $H$, $K$, and $F560W$ in our analysis, increase the existing optical data, and extend the observations up to nearly 60 days. We denote this set as \textbf{``Extended''}. Both sets are corrected for extinction using the SF11 assumption for the foreground extinction, see.~\S\ref{SED}. We made the choice not to take into account any radio data, as only a subset was publicly available at the time of the publication of the article (e.g. data from \citealt{2023Laskar} was not yet available). 
For these two multi-wavelength data sets and for both Bayesian Inference presented below, we use the same parameter space and priors, presented in table \ref{tab:MCMC_paramsNMMA}, except for the initial Lorentz factor $\Gamma_0$, 
which is needed only in the second model including the coasting phase.
%which is needed to describe the coasting phase in the IAP models (and is not included in \texttt{NMMA}).}
The luminosity distance to the source is fixed to $D_\mathrm{L} = 742~\mathrm{Mpc}$, corresponding to a redshift $z=0.151$ for a flat cosmology with $H_0 = 67.7~\mathrm{km}\cdot\mathrm{s}^{-1}\cdot\mathrm{Mpc}^{-1}$ and $\Omega_\mathrm{m}=0.307$ \citep{2016A&A...594A..13P}.

\subsubsection{Bayesian Inference using \texttt{NMMA}, investigation of the jet structure and SN contribution}
\label{NMMA}

As a further framework to interpret GRB 221009A, we use the Nuclear physics and Multi-Messenger Astronomy framework \texttt{NMMA}~\citep{Dietrich:2020efo,Pang:2022rzc}\footnote{\url{https://github.com/nuclear-multimessenger-astronomy/nmma}} that allows us to perform joint Bayesian inference of multi-messenger events containing gravitational waves, kilonovae, SNe, and GRB afterglows. We have analyzed both the \textit{GRANDMA} and \textit{Extended} data sets with \texttt{NMMA}.

For this work, we follow~\cite{2023arXiv230102049K} and employ first the top-hat jet structure (with on axis-assumption and with a free case) with the semi-analytic code \texttt{afterglowpy} \citep{vanEerten:2009pa,Ryan:2019fhz}.\footnote{The nested sampling algorithm implemented in \textsc{pymultinest} \citep{2016ascl.soft06005B} is used.}
In this model, the dynamics of the relativistic ejecta propagating through the interstellar medium are treated under the thin-shell approximation, and the angular structure is introduced by dissecting the blast wave into angular elements, each of which is evolved independently, including lateral expansion. Magnetic-field amplification, electron acceleration, and the synchrotron emission from the forward shock are treated according to the analytical prescriptions of~\citet{Sari:1997qe}. The observed radiation is computed by performing equal-time arrival surface integration. It is important to note that the model does not account for the presence of the reverse shock or the early coasting phase and does not include inverse Compton radiation. This limits its applicability to the early afterglow of very bright GRBs. 

While we find a more steeply decaying emission component at early times, it is unclear whether it can be attributed to the reverse shock emission \citealt{2023Laskar}.

The advantage of the NMMA framework is the possibility of comparing different astrophysical scenarios and models in a straightforward way. 
As a starting point, we compare different jet structures. In addition to the top-hat jet, 
we also employed a Gaussian and Power-law jet. Gaussian jet features an angular dependence $E(\theta_{\rm obs}) \propto \exp(-\theta_{\rm obs}^2/(2\theta_c^2))$ for $\theta_{\rm obs}\leq \theta_w$, with $\theta_w$ being an additional free parameter. The power-law jet features an angular dependence $E(\theta_{\rm obs}) \propto (1 + (\theta_{\rm obs} / \theta_c)^2 / b)^{-b/2}$ for $\theta_{\rm obs} \leq \theta_w$, with $\theta_w$ and $b$ being an additional parameter.

\begin{figure}[t]
\centering
\includegraphics[width=\columnwidth]{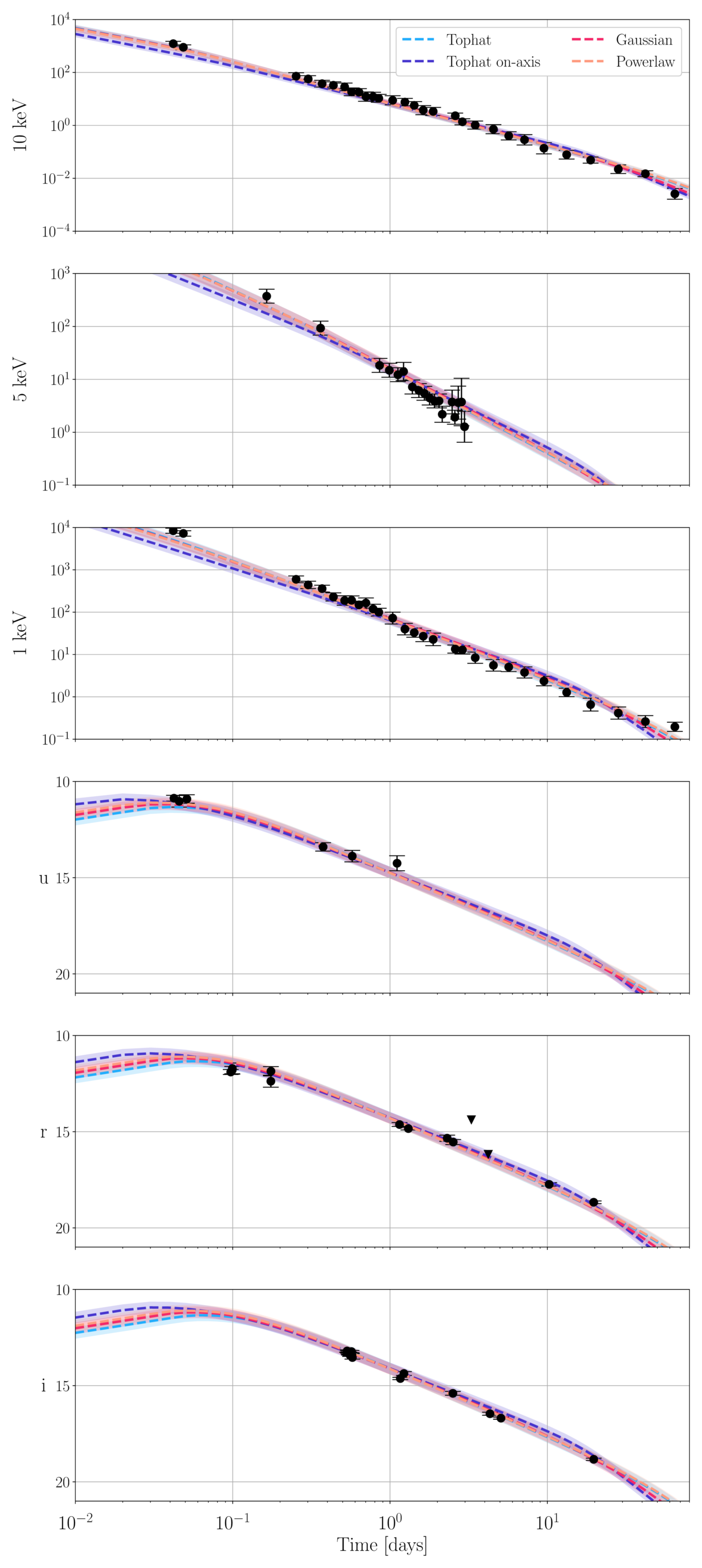}
  \caption{ \texttt{NMMA} - Observational data (\textit{GRANDMA} data) and best-fit light curves of selected filters for the NMMA analysis using the SF11 extinction and the four employed jet structures. The X-ray bands are shown in $\mu$Jy and the rest of the bands are shown in AB magnitude.}
   \label{fig:NMMA_light curve_GRANDMA}
\end{figure}

We present our best-fit light curves for the SF11 extinction, with different jet structures assumed in Fig.~\ref{fig:NMMA_light curve_GRANDMA}, which shows the \textit{GRANDMA} data (see Fig.~\ref{fig:NMMA_light curve} in the Appendix for the \textit{Extended} data).

%\textcolor{red}{It's a bit late for that but wouldn't it be better to compare XRT only vs HXMT only rather than XRT only vs XRT + HXMT ? In the latter you still expect the XRT data to constrain the model a lot, even when including HXMT data.} \textbf{[ALEX: Yes, a bit late...]} \textbf{[Peter: we also have that run done, so if u want such a comparison]}
We find that the observational data are only moderately well-fit by the model. While the $r$-band lightcurve is reasonably well-recovered, the predicted lightcurves at higher frequencies, especially in X--rays, cannot reproduce the observed evolution.

%Computing the log Bayes factor 
%\begin{equation}
% \ln \mathcal{B}^{\rm SF11}_{\rm RF09} =  \ln \frac{p(d | {\rm GRB_{\rm model}, \rm SF11})} {p(d | {\rm GRB_{\rm model}, \rm RF09)}}
%\end{equation}
%for our two extinctions of SF11 vs.~RF09, we obtain $0.238\pm0.095$ when we do not include HXMT data and a log Bayes factor of SF11 against RF09 of $0.325\pm 0.097$ if HXMT data are included, i.e., following Jeffrey's scale, we find that this difference is not statistically significant. \\

We present the corresponding source parameters, namely, the inclination angle $\theta_{\rm obs}$, isotropic energy $E_0$, the interstellar medium density $n_{\rm ism}$, half-opening angle of the jet core $\theta_{\rm core}$, and microphysical parameters $\{p, \epsilon_e, \epsilon_B, \zeta\}$ (the power-law index of the electron energy distribution, the fraction of energy in electrons, the fraction of energy in the magnetic field and the fraction of electron accelerated, respectively) using the four different jet structure models in Fig.~\ref{fig:NMMA_corner_GRANDMA}, which uses the \textit{GRANDMA} data (see Fig.~\ref{fig:NMMA_corner} in the Appendix for the \textit{Extended} data); each simulation uses $2048$ live points for the nested sampling. The full posteriors can be found in Tab.~\ref{tab:MCMC_paramsNMMA}.

%In general, we find consistent results within statistical uncertainties (quoted and shown at the 90\% credible level) larger than the differences caused by the input data when analyzing the four data sets. 
Most surprising in our analysis might be the relatively large jet-opening angle (the viewing angle being near the edge but still within the jet), which might be hard to explain given the high isotropic energy release of the GRB. For the tophat on-axis (i.e. $\theta_{\rm obs} = 0$), the light curve seems dimmer than expected for the X-ray data, which drives the analysis to prefer larger viewing angles. %However, this early $u$-band data may contain a contribution from another emitting component, as discussed above. 
%\textcolor{red}{This sentence has to be removed or largely rewritten}
%Nevertheless, considering that the lower bound of the posterior does reach values of a few degrees and that the inclusion of the $0.3$ magnitude uncertainty generally leads to wider, less restrictive posteriors, we find consistent results with other analyses performed in this article and also previously shown in the literature, e.g., \cite{2023Laskar}.
Although a larger viewing is preferred, the relation of $\theta_{\rm obs} < \theta_{\rm core}$ is clearly observed in the posterior of the angle ratio $\theta_{\rm obs} / \theta_{\rm core}$ (in Tab.~\ref{tab:MCMC_paramsNMMA}), which is attributed to the absence of the jet break in the present data. 
We also provide the $\chi^2 / {\rm d.o.f.}$ for these fits in Tab.~\ref{tab:NMMA_jet_chi2}

\begin{figure*}[t]
\centering
\includegraphics[width=\textwidth]{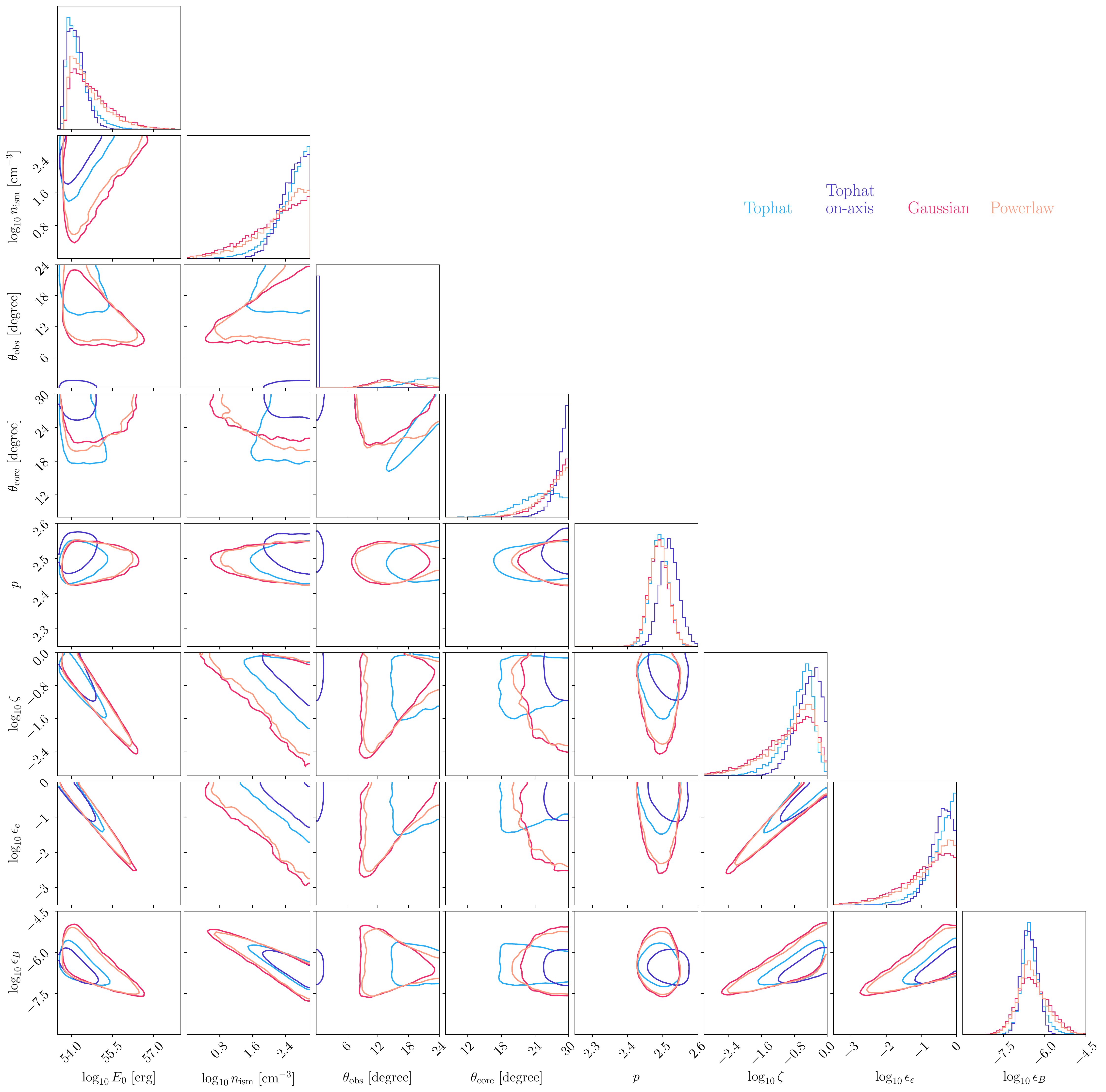}
  \caption{ \texttt{NMMA} - Posterior distribution (shown are 90\% confidence intervals) for our selected data sets when using different jet models of \texttt{afterglowpy} analyzing the \textit{GRANDMA} data.}
  \label{fig:NMMA_corner_GRANDMA}
\end{figure*}

\begin{table*}
%\begin{sidewaystable}
	\renewcommand{\arraystretch}{1.2}
	\centering
	\caption{ \texttt{NMMA} - Parameters and prior bounds employed in our Bayesian inferences. We report median posterior values at 90~\% credibility from simulations that were run with different jet structures using SF11 extinction data for analysis (\textit{GRANDMA} and \textit{Extended}), see \S\ref{NMMA} and Figs.~\ref{fig:NMMA_light curve},\ref{fig:NMMA_corner}.}
	\label{tab:MCMC_paramsNMMA}
     %\begin{turn}{90}
     \resizebox{0.65\paperheight}{!}{
	\begin{tabular}{l c l  c c c c c c c c}
	\hline
	Parameter & Bounds & Prior & \multicolumn{2}{c}{Tophat} & \multicolumn{2}{c}{Tophat on-axis} & \multicolumn{2}{c}{Gaussian} & \multicolumn{2}{c}{Power-law} \\
     &  &  & \textit{GRANDMA} & \textit{Extended} & \textit{GRANDMA} & \textit{Extended}  & \textit{GRANDMA} & \textit{Extended} & \textit{GRANDMA} & \textit{Extended} \\
	\hline
	Isotropic afterglow energy $E_0$ [erg]& [$10^{50}$,$10^{58}$] & log-uniform & $10^{54.16^{+0.71}_{-0.45}}$ & $10^{54.71^{+0.91}_{-0.80}}$ & $10^{+54.15^{+0.52}_{-0.45}}$ & $10^{54.51^{+0.52}_{-0.57}}$ & $10^{54.62^{+1.18}_{-0.81}}$ & $10^{55.27^{+1.20}_{-1.10}}$ & $10^{54.17^{+1.15}_{-0.68}}$ & $10^{55.13^{+1.22}_{-1.05}}$\\
	Ambient medium's density $n_{\rm ism} $[$\rm{cm}^{-3}$]& [$10^{-6}$,$10^{3}$] & log-uniform & $10^{2.61^{+0.39}_{-0.70}}$ & $10^{2.60^{+0.40}_{-0.63}}$ & $10^{2.61^{+0.39}_{-0.49}}$ & $10^{2.51^{+0.49}_{-0.51}}$ & $10^{2.27^{+0.73}_{-1.14}}$ & $10^{2.48^{+0.52}_{-0.82}}$ & $10^{+2.39^{+0.61}_{-1.04}}$ & $10^{2.51^{+0.49}_{-0.76}}$\\
	%Thermal
    Energy fraction in electrons $\epsilon_{\mathrm{e}}$ & [$10^{-4}$,$1$] & log-uniform & $10^{-0.38^{+0.38}_{-0.70}}$ & $10^{-0.96^{+0.66}_{-0.81}}$ & $10^{-0.41^{+0.41}_{-0.47}}$ & $10^{-0.68^{+0.49}_{-0.51}}$ & $10^{-0.85^{+0.78}_{-1.09}}$ & $10^{-1.39^{+0.98}_{-1.08}}$ & $10^{-0.75^{+0.67}_{-1.07}}$ & $10^{-1.27^{+0.93}_{-1.09}}$\\
	%Thermal
    Energy fraction in magnetic field $\epsilon_{\mathrm{B}}$ &  [$10^{-9}$,$1$] & log-uniform & $10^{-6.54^{+0.62}_{-0.49}}$ & $10^{-6.59^{+0.64}_{-0.56}}$ & $10^{-6.56^{+0.49}_{-0.42}}$ & $10^{-6.83^{+0.55}_{-0.49}}$ & $10^{-6.46^{+1.02}_{-0.86}}$ & $10^{-6.71^{+0.84}_{-0.71}}$ & $10^{-6.50^{+0.90}_{-0.80}}$ & $10^{-6.69^{+0.79}_{-0.70}}$\\
	Electron distribution power-law index $p$ & [$2$,$3$] & uniform & $2.49^{+0.04}_{-0.04}$ & $2.39^{+0.03}_{-0.03}$ & $2.52^{+0.04}_{-0.04}$ & $2.53^{+0.03}_{-0.03}$ & $2.49^{+0.04}_{-0.05}$ & $2.40^{+0.03}_{-0.03}$ & $2.49^{+0.04}_{-0.04}$ & $2.40^{+0.03}_{-0.03}$\\
        Fraction of accelerated electrons $\zeta$ & [$10^{-4}$,$1$] & log-uniform & $10^{-0.63^{+0.45}_{-0.67}}$ & $10^{-0.39^{+0.39}_{-0.62}}$ & $10^{-0.44^{+0.44}_{-0.47}}$ & $10^{-0.47^{+0.47}_{-0.52}}$ & $10^{-0.85^{+0.78}_{-1.09}}$ & $10^{-0.52^{+0.52}_{-0.83}}$ & $10^{-0.75^{+0.67}_{-1.07}}$ & $10^{-0.47^{+0.47}_{-0.80}}$\\ 
        Viewing angle $\theta_{\rm obs}$ [degrees] & [$0$,$30$] & uniform & $21.45^{+4.79}_{-5.56}$ & $17.14^{+4.82}_{-4.50}$ & $0$ & $0$ & $13.98^{+5.75}_{-5.54}$ & $12.32^{+4.87}_{-5.04}$ & $15.36^{+7.33}_{-6.31}$ & $13.19^{+5.34}_{-5.45}$\\
	Jet core's opening angle $\theta_{\rm core}$ [degrees] &  [$0.1$,$30$] & uniform & $24.91^{+5.09}_{-5.50}$ & $23.68^{+6.31}_{-5.42}$ & $28.85^{+1.15}_{-2.37}$ & $29.58^{+0.42}_{-0.98}$ & $27.65^{+2.35}_{-4.17}$ & $27.85^{+2.15}_{-4.02}$ & $27.29^{+2.71}_{-4.80}$ & $27.56^{+2.44}_{-4.00}$\\
        ``Wing'' truncation angle $\theta_{\rm wing}$ [degrees] & [$0.1$,$30$] & uniform &  \multicolumn{2}{c}{$-$} & \multicolumn{2}{c}{$-$} & $16.65^{+6.81}_{-6.83}$ & $17.51^{+7.01}_{-7.19}$ & $18.33^{+8.99}_{-7.17}$ & $18.70^{+7.49}_{-7.54}$ \\
        Power-law structure index $b$ & [0.1, 7] & uniform &  \multicolumn{2}{c}{$-$} & \multicolumn{2}{c}{$-$} & \multicolumn{2}{c}{$-$} & $2.42^{+3.62}_{-2.32}$ & $2.64^{+3.41}_{-2.54}$\\
        Angle ratio $\theta_{\rm obs} / \theta_{\rm core}$ & [$1/300$,$300$] & - & $0.866^{+0.05}_{-0.06}$ & $0.725^{+0.04}_{-0.04}$ & $0$ & $0$ & $0.515^{+0.212}_{-0.189}$ & $0.454^{+0.165}_{-0.192}$ & $0.573^{+0.304}_{-0.232}$ & $0.491^{+0.180}_{-0.218}$ \\
        \hline
	\end{tabular}
        }
      %  \end{turn}
 %\end{sidewaystable}
\end{table*}

\begin{table}[]
    \centering
     \resizebox{1.0\columnwidth}{!}{
    \begin{tabular}{|l|c c|}
    \hline
         & \textit{Extended} data & \textit{GRANDMA} data\\
    \hline
      Tophat $\theta_{\rm obs} \leq 0$ vs $\theta_{\rm obs} = 0$  & $71.26\pm0.14$ & $30.66\pm0.13$ \\
      Tophat vs Gaussian & $4.29\pm0.14$ & $3.86\pm0.14$ \\
      Tophat vs Power-law & $3.53\pm0.14$ & $3.19\pm0.14$\\
    \hline
    \end{tabular}}
    \caption{The log Bayes factor between different models. For both the \textit{Extended} data and \textit{GRANDMA} data, the tophat model without on-axis assumption is preferred.}
    \label{tab:NMMA_jet_logB}
\end{table}

\begin{table}[]
    \centering
    \begin{tabular}{|l|c c|}
    \hline
         & \textit{Extended} data & \textit{GRANDMA} data\\
    \hline
      Tophat $\theta_{\rm obs} \leq 0$ & $0.551$ & $0.496$ \\
      Tophat $\theta_{\rm obs} = 0$ & $0.882$ & $0.972$ \\
      Gaussian & $0.561$ & $0.542$ \\
      Power-law & $0.555$ & $0.520$ \\
    \hline
    \end{tabular}
    \caption{The $\chi^2 / {\rm d.o.f.}$ of different models for \texttt{NMMA}.}
    \label{tab:NMMA_jet_chi2}
\end{table}

%In addition to the top-hat jet for which we showed results before, 
%we also employed a Gaussian jet, which features an angular dependence $E(\theta_{\rm obs}) \propto \exp(-\theta_{\rm obs}^2/(2\theta_c^2))$ for $\theta_{\rm obs}\leq \theta_w$, with $\theta_w$ being an additional free parameter and a power-law jet in which the energy scales as $E(\theta_{\rm obs}) \propto (1 + (\theta_{\rm obs} / \theta_c)^2 / b)^{-b/2}$ for $\theta_{\rm obs} \leq \theta_w$, with $b$ being an additional parameter. %For simplicity, we restrict this comparison to the full \textit{Swift} and HXMT data set and the SF11 extinction.

%Overall, we do not find strong statistical evidence for the presence of jet structure and the top-hat jet model remains preferred (due to Occam's razor penalty to more complicated models). 

Finally, following the study of \citet{2023arXiv230111170F}, we investigate the possibility of an SN connected to GRB 221009A. For modeling the SN, we use the \texttt{nugent-hyper} model from \texttt{sncosmo}~\citep{Levan:2004sn} with a shift in the absolute magnitude, $S_{\rm{max}}$, as the main free parameter. We vary this free parameter within $S_{\rm max}\in[-30\;\rm mag,\;30\;\rm mag]$. 
The nugent-hyper model is a template constructed from observations of SN 1998bw \citep{1998Natur.395..670G}. Within our analysis, we find that in our runs combining the GRB top-hat jet afterglow with a SN component, the $\chi^2 / {\rm d.o.f.}$ are very similar to the pure top-hat jet model ($0.551$ to $0.548$ for \textit{Extended} data and $0.496$ to $0.503$ for \textit{GRANDMA} data). Due to the size of the parameter space, the log Bayes Factor prefers the simpler top-hat jet model, $0.542\pm0.140$ for the \textit{Extended} data and $0.540\pm0.136$ for the \textit{GRANDMA} data). Hence, there is no strong evidence for or against the presence of a SN contribution consistent with the nugent-hyper model combined with models used in our analysis (see, for instance, \cite{2023arXiv230111170F} for another interpretation. 

%\subsubsection{\cp{Results interpretation using the IAP model}}
\subsubsection{Refining the physics in the top-hat jet model}
\label{IAP}

The poor quality of the fits obtained in \S\ref{NMMA} is a clear indication of tension between the observed temporal and spectral slopes, as suggested by other authors, e.g. \cite{2023Laskar, 2023JHEAp..37...51S,2023arXiv230207906O}. We also note that some of the parameters obtained and presented in Figs. \ref{fig:NMMA_corner} and \ref{fig:NMMA_corner_GRANDMA} are at odds with typical GRB afterglow parameters.
%\cp{Some of the parameters obtained in \S\ref{NMMA} and presented in Figs. \ref{fig:NMMA_corner} and \ref{fig:NMMA_corner_GRANDMA} are at odds with typical GRB afterglow parameters.}
Perhaps the most striking is the very large jet opening angle. It is constrained to $\theta_\mathrm{core} \gtrsim 15$degrees, while typical jets have opening angles $\theta_\mathrm{core} \simeq (2.5\pm1.0)$degree \citep{2018ApJ...859..160W}. 

The \texttt{afterglowpy} model, which is used by \texttt{NMMA}, despite being already a refined implementation of the external shock afterglow model \citep{Ryan:2019fhz}, suffers from some limitations.
%\cp{Because \texttt{NMMA} uses \texttt{afterglowpy} to model the GRB afterglows, there are some intrinsic limitations to the model. 
The dynamics of the jet deceleration is assumed to be in the self-similar regime at all times, which can lead to flux over-estimates at very early times when the jet still is in the coasting phase. We also note that Synchrotron Self-Compton (SSC) scattering is not accounted for; this could be important in the case of GRB 221009A, where some very high energy photons observed may hint towards strong SSC emission.
To further validate the previous analysis, we therefore also model the afterglow data of GRB 221009A using the afterglow model from \citet{pellouin} which
not only includes synchrotron radiation but also computes the 
%another model developed at IAP assuming synchrotron and 
Synchrotron Self-Compton (SSC) radiation, taking into account both the Thomson and Klein-Nishina regime with a treatment following \citet{2009ApJ...703..675N}.
%at the forward shock of a relativistic blast wave propagating through the ISM. 
%In this model, SSC diffusions can occur in both Thomson and Klein-Nishina regimes and are treated as first introduced in \citet{2009ApJ...703..675N}.
This model also accounts for the jet lateral structure, any viewing angle, and %, although in this case we do not include any lateral structure and fix the viewing angle $\theta_\mathrm{obs} = 0\degree$
%We also assume a constant-density ISM. %Another difference with \texttt{afterglowpy} used in \texttt{NMMA} is that 
%This model 
also includes the treatment of the coasting phase of the jet propagation, which can induce differences at early times.
A detailed %model 
description of this afterglow model will be provided in \cite{pellouin}. 
However, an analysis with the best-fit parameters shows it does not impact the light curves post the first observed time at 3618\,s by \textit{Swift}-XRTin this case. 

We used a Markov Chain Monte Carlo (MCMC) routine to infer the physical parameters for the afterglow, %using the optical data with SF11 extinction, as well as the HXMT X-ray data between 1.5 and 10 keV. For the $\chi^2$ computation, we slightly modify the errors to avoid any over-fitting of points with very small errors, using $\max \{\textrm{flux error}; 0.3 \times \textrm{flux}\}$. All data points are jointly fitted. 
using the data set presented at the beginning of \S\ref{sec:modelling}. When performing the $\chi^2$ computation, we inflate the errors to avoid any over-fitting of points with artificially small errors, using $\max \{\textrm{flux error}; 0.3 \times \textrm{flux}\}$.
We initialize $100$ independent chains, and run them over $20000$ iterations; we remove chains that get stuck in a high $\chi^2$ region of the parameter space, as they are not true solutions.

\begin{figure*}%[!th]
\centering
\includegraphics[width=\textwidth]{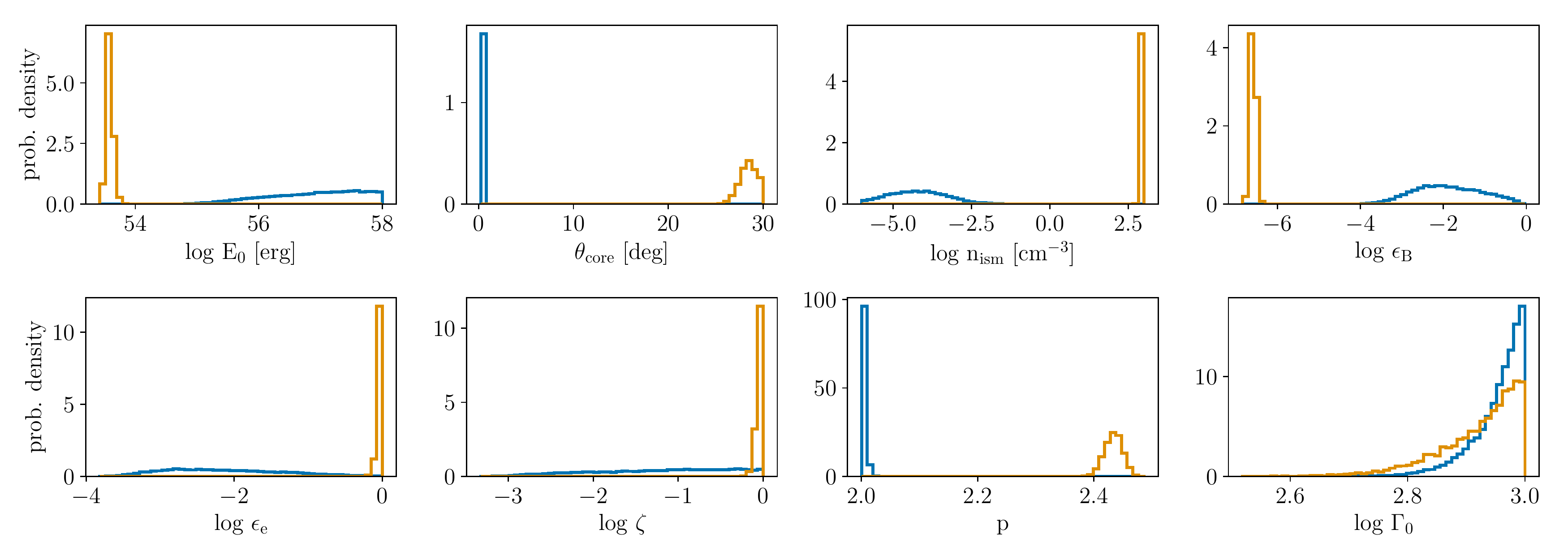}
  \caption{Bayesian inference presented in \S\ref{IAP}: Marginalized posterior distribution of the free parameters best fitting the extended data with an on-axis observation of a top-hat jet radiating via synchrotron only, as presented in \S\ref{IAP}. In blue, marginalized distributions for the model which features a low $\theta_\mathrm{core}$ and $p \sim 2$, not found with \texttt{NMMA}. In orange, marginalized distributions for the model with a high $\theta_\mathrm{core}$, similar to what is found with \texttt{NMMA}. The median values and 90~\% confidence intervals can be found in Table~\ref{table:MCMC_IAP_posterior}.}
   \label{fig:IAP_MCMC_marginalized_posterior}
\end{figure*}

\begin{figure*}%[!th]
\centering
\includegraphics[width=\textwidth]{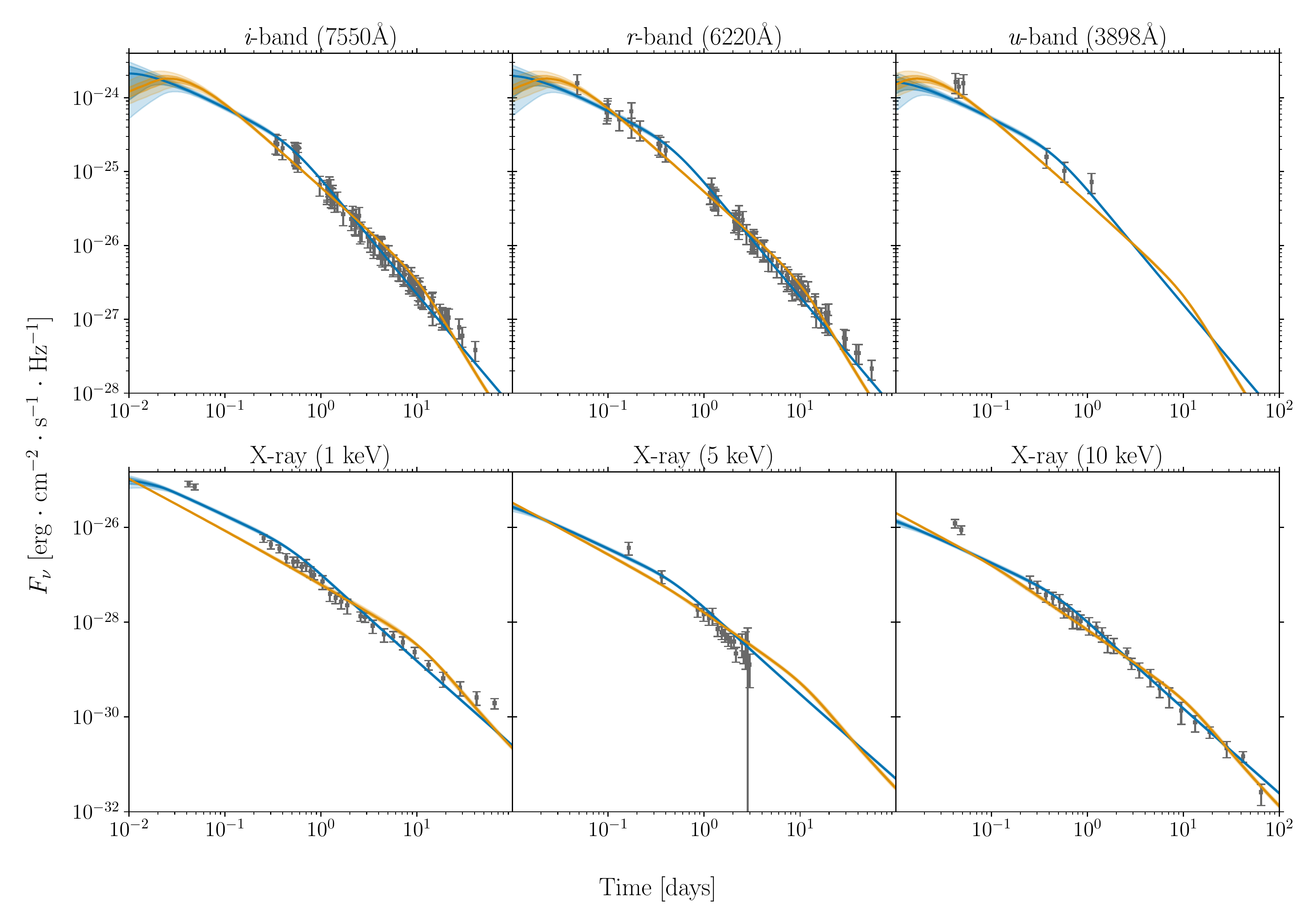}
  \caption{Bayesian inference presented in \S~\ref{IAP}}
  - Predicted light curves for the two classes of parameters reported in Fig. \ref{fig:IAP_MCMC_marginalized_posterior} and found using a top-hat model with a fixed observing angle $\theta_\mathrm{obs} = 0$degree, and assuming only synchrotron radiation. Observing frequencies or energies are shown on top of each panel, and the fitted observational data is displayed in grey. Blue curves show the model with a low $\theta_\mathrm{core}$ and $p \sim 2$. Orange curves show the model with a high $\theta_\mathrm{core}$.
   \label{fig:IAP_best_fit_models}
\end{figure*}

Our first analysis uses a simplified model where only synchrotron radiation powers the afterglow emission, for comparison with the analysis presented in \S\ref{NMMA}. For the top-hat jet with a fixed viewing angle $\theta_\mathrm{obs} = 0$degree, the posterior samples converge towards parameter values that are very similar to those presented in \S\ref{NMMA}, both when fitting the GRANDMA data and the extended data. Those values are listed in Table \ref{table:MCMC_IAP_posterior}, and the results are presented in Fig. \ref{fig:IAP_best_fit_models}. Our analysis with this independent model confirms the relatively poor quality of the best fit.
However, when fitting the extended data set, another solution emerges in the MCMC posteriors with this model; this fit has a lower $\chi^2$, but still leads to a best fit of poor quality. Contrary to \texttt{NMMA} using \texttt{afterglowpy}, the model does not include the lateral expansion of the jet at late times, which leads to an alternative scenario with an early jet break with a shallow post-jet break decay slope. This solution %favors 
implies a small core jet opening angle  $\theta_\mathrm{core} \lesssim 0.7$degree
and an electron slope very close to $2$. 
%very low core jet opening angles $\theta_\mathrm{core} \lesssim 0.7\degree$. 
Both solutions to the model appear on an equal number of MCMC chains (44 each) and therefore correspond to 2 local minima for the $\chi^2$ distribution, with %equal 
similar weight. When isolating both solutions, we observe the usual parameter correlations (e.g. $E_\mathrm{0,iso}$ and $n_\mathrm{ism}$ or between $\epsilon_\mathrm{B}$, $\epsilon_\mathrm{e}$ and $\zeta$). For better readability, we show in Fig. \ref{fig:IAP_MCMC_marginalized_posterior} only the marginalized posterior distribution for the 8 free parameters of the model, and split the two classes of models in two colors. In blue, we show marginalized posterior distributions for this new model (with a low $\theta_\mathrm{core}$ and $p \lesssim 2.02$) and in orange, marginalized posterior distributions for the model with a large $\theta_\mathrm{core}$, similar to what is found in \S\ref{NMMA}. The median values and the 90~\% confidence intervals are reported in Table~\ref{table:MCMC_IAP_posterior} for the two models.

\begin{table*}
	\renewcommand{\arraystretch}{1.2}
	\centering
	\caption{
 Bayesian inference presented in \S\ref{IAP}: Parameters and prior bounds employed in our Bayesian inferences. We report median posterior values at 90~\% credibility from simulations that were run with a top-hat jet structure with a fixed observing angle $\theta_\mathrm{obs} = 0$degrees assuming synchrotron radiation. We fit the extended data set presented in \S\ref{sec:modelling}. These results are discussed in \S\ref{IAP}, and in Figs. \ref{fig:IAP_MCMC_marginalized_posterior}, \ref{fig:IAP_best_fit_models}.}
	\begin{tabular}{l c c l c c}
	\hline
	Parameter & Symbol & Bounds & Prior & \multicolumn{2}{c}{Posterior}\\
 	 &  &  &  & low $\theta_\mathrm{core}$ & high $\theta_\mathrm{core}$\\
	\hline

	Isotropic afterglow energy [erg]& $E_{0}$ & [$10^{50}$,$10^{58}$] & log-uniform & $10^{57.01^{+0.99}_{-1.16}}$ & $10^{53.58^{+0.09}_{-0.08}}$\\
 	Opening angle of the core of the jet [deg] & $\theta_{\mathrm{core}}$ &  [$0.1$,$30$] & uniform & $0.39^{+0.13}_{-0.11}$ & $28.47^{+1.52}_{-1.18}$\\
	Density of the ambient medium [$\rm{cm}^{-3}$]& $n_{\mathrm{ism}}$ &  [$10^{-6}$,$10^{3}$] & log-uniform & $10^{-4.23^{+1.36}_{-1.51}}$ & $10^{2.98^{+0.02}_{-0.04}}$\\
 	Fraction of the energy which generates the magnetic field & $\epsilon_{\mathrm{B}}$ &  [$10^{-9}$,$1$] & log-uniform & $10^{-1.93^{+1.39}_{-1.17}}$ & $10^{-6.59^{+0.11}_{-0.10}}$\\
	Fraction of the energy which accelerates the electrons & $\epsilon_{\mathrm{e}}$ & [$10^{-4}$,$1$] & log-uniform & $10^{-2.22^{+1.33}_{-1.16}}$ & $10^{-0.02^{+0.02}_{-0.05}}$\\
    Fraction of electrons accelerated at the shock & $\zeta$ & [$10^{-4}$,$1$] & log-uniform & $10^{-1.10^{+1.10}_{-1.25}}$ & $10^{-0.04^{+0.04}_{-0.06}}$\\
	Electron population Lorentz factor injection index & $p$ &  [$2$,$3$] & uniform & $2.003^{+0.005}_{-0.003}$ & $2.43^{+0.03}_{-0.02}$\\
 	Initial Lorentz factor & $\Gamma_{0}$ &  [$10^{1}$,$10^{3}$] & log-uniform & $10^{2.96^{+0.04}_{-0.07}}$ & $10^{2.94^{+0.06}_{-0.11}}$\\
    \hline
	\end{tabular}
 \label{table:MCMC_IAP_posterior}
\end{table*}

We investigate these two types of models and show in Fig. \ref{fig:IAP_best_fit_models} light curves obtained with the two posterior samples of parameters. The light curves are computed in 3 optical/UV bands (\textit{i}-band, $755~\mathrm{nm}$, \textit{r}-band, $622~\mathrm{nm}$, \textit{u}-band, $389.8~\mathrm{nm}$) and in the 3 X-ray bands (1 keV and 10 keV, with XRT observations; 5 keV with \textit{HXMT} observations). All the fitted observations are shown in grey.

Using Fig. \ref{fig:IAP_best_fit_models}, we see that though both classes of models are able to broadly reproduce the multi-wavelength observations, they both fail to reproduce the observed features accurately. The models with high $\theta_\mathrm{core}$, also found with \texttt{NMMA} and shown in orange, accurately fit the optical but fail to reproduce the observed X--ray temporal slope. They also feature a jet break at $t \sim 8~\mathrm{days}$, which is not observed in the data. Conversely, the models with low $\theta_\mathrm{core}$, though slightly favored statistically ($\chi^2 = 600$ for the latter model, $\chi^2 = 950$ for the former with the same d.o.f.), also pose major interpretation challenges. The temporal decay in X-rays is also not in line with X-ray observations, and the very high $E_0$ values that are found (between $10^{55}$ and $10^{58}$ ergs) imply a very low prompt efficiency, given a prompt energy $E_{\gamma, \mathrm{iso}} \sim 10^{55}~\mathrm{erg}$ \citep{2023AnHXMT}. 

While the true energy in the jet ranges between $10^{51}$ and $5\times 10^{53}~\mathrm{erg}$ due to the very narrow jet opening angles $\theta_\mathrm{core} \lesssim 0.7$degree, the posterior distribution of the prompt efficiency peaks at $10^{-4}$, which seems very low for most prompt emission models, especially considering the bright luminosity of the prompt emission of GRB 221009A. This posterior distribution shows, however, a tail towards larger values, reaching $10^{-1}$, which are more physically plausible. the physical relevance of this scenario is also questionable regarding the jet dynamics, as at least a moderate lateral expansion should be expected.

As a final verification, we also computed the predicted light curves at radio frequencies and compared them with observations reported in \citet{2023Laskar} and \citet{2023arXiv230207906O}. With both classes of models, our predictions over-estimate the flux compared to the observations.
We therefore, conclude that while these parameters do correspond to the best fits of the extended data set using a top-hat jet model with a fixed observing angle $\theta_\mathrm{obs} = 0$degree and assuming only synchrotron radiation, they do not provide satisfactory predictions given other constraints found in the literature, motivating a deeper analysis.
When accounting for SSC scattering in the model, we ran the models with narrow $\theta_\mathrm{core}$. We also tested models with a free observing angle and find similar results to those presented in \S\ref{NMMA}: this model is preferred, but the typical values found for the parameters are close to those found with the fixed observation angle. Therefore our analysis shows that even in a more realistic description including the coasting phase and the SSC radiation, the standard top-hat jet afterglow model is in tension with the observed data.
We note that for free observing angles, the MCMC chains do not favor narrow $\theta_\mathrm{core}$ anymore. Finally, we also study the impact of the jet structure: our findings are similar to those presented in \S\ref{NMMA}. We also performed our analysis with only the GRANDMA data set and find similar values as those found with \texttt{NMMA}, but in this case the MCMC chains also do not find the very narrow $\theta_\mathrm{core}$ models. 
We leave to a future study the investigation of more advanced models regarding, for instance, the jet structure of the external medium density.

\section{Discussion and Conclusion}
\label{conclusion}

In this work, the properties of the GRB 221009A afterglow are studied using a multi-wavelength data set, presenting data from optical observations from ground-based telescopes or the GRANDMA/KNC network and the Low Energy X-ray telescope (LE) onboard the \textit{Insight}-\textit{HXMT} satellite. The X-ray observations were made 9.8 hours to 3 days after the trigger time, while the ultraviolet, optical, and near-infrared sky was covered from the prompt emission (shallow limits from all-sky cameras) and then (with narrow-field instruments) from 2.2 hours after the trigger time to about 20 days. The GRANDMA network involved more than 30 telescopes, including both professional and amateur telescopes, and collected more than 200 images for this GRB. This is one of the few GRB afterglows that has been observed extensively by amateur astronomers. The measurements with the deepest limiting magnitudes reach $m_{lim}=24.6$ mag in $g^\prime$ band by a professional telescope (CFHT) and $m_{lim}=21.5$ mag in the $V$ band by an amateur telescope, demonstrating the potential for citizen contributions to time-domain astrophysical science. We also collected prompt observations of the GRB in the optical (between $T_0$ and to $T_0$+500 seconds) by cameras managed by the Desert Fireball Network, but no optical flash was detected in the $V$ band (down to a limiting magnitude of 3.8 mag). We furthermore collect public data from the XRT telescope onboard the \textit{Swift} satellite, with the first observation having been taken about one hour after the GRB trigger time. Two specially-tuned photometric pipelines, {\sc STDPipe} and {\sc Muphoten}, are used to analyze the GRB afterglow data. The observations are calibrated using stars from the PS1 catalog; slightly different results being obtained for Johnson-Cousins filters between the two pipelines. For this reason, only a subset of data with good quality and consistent results have been selected for analysis. 

In this paper, we tackle the challenge of determining the significant extinction correction, as the GRB lies behind the Galactic plane. To correct for this, we employ two different techniques: firstly, we use the SF11 maps \citep{Schlafly2011}, which may overestimate the extinction. Secondly, we use the RF09 maps by \citet{2009MNRAS.395.1640R}, which utilize near-infrared color excess determinations based on 2MASS observations. This method results in a significantly lower extinction value (but is only valid out to $2-3$ kpc), compared to the SF11 value. Taking into account the existence of dust at larger distances determined by X-ray measurements of dust rings \citep{2023arXiv230101798N,Vasilopoulos2023,2023Swift}, we proceed to discuss the reliability of these measurements and conduct our follow-up analysis using both correction methods for comparison.

%This method results in a significantly lower extinction value of $A_V=2.195$/$E_{(B-V)}=0.709$ mag (but is only valid out to $2-3$ kpc), compared to the SF11 value of $A_V=4.1$/$E_{(B-V)}=1.32$ mag. Taking into account the existence of dust at larger distances determined by X-ray measurements of dust rings \citep{2023arXiv230101798N,Vasilopoulos2023,2023Swift}, we proceed to discuss the reliability of these measurements and conduct our follow-up analysis using both correction methods for comparison.

Empirical analysis of the light curve shows it to be composed of three power-law sections (steep, shallow, steep), with the first section only covered by a short data baseline. The data after $\sim0.1$ d show a clear break and a relatively shallow post-break slope with no further indication of a jet break, which would usually lead to a decay slope $\alpha\gtrsim2$. The light curve analysis yields a SED which we fit with three solutions for the foreground/host-galaxy extinction including one under the assumption that the entire extinction is foreground. All extinction models yield viable solutions; in combination with spectroscopic evidence for small-to-moderate host-galaxy extinction, we prefer the combination of SF11 foreground correction and about half a magnitude of SMC-type host-galaxy extinction. Using these values, we are able to compare the optical afterglow to a global sample and find it to be luminous but not excessively so, in contrast to the extreme isotropic energy release of the prompt emission, a result also found for the X-ray afterglow \citep{2023Swift}.

We analysed our observations in the framework of the standard GRB afterglow model; in this model, the observed flux is dominated by synchrotron radiation from shock-accelerated electrons at the forward external shock due to the deceleration of the GRB relativistic jet by the ambient medium \citep{Sari:1997qe}. We limited our study to the case of a uniform medium. We performed Bayesian Inference using two multi-wavelength datasets, the first composed of our own GRANDMA data, complemented by X-ray data (\textit{Swift}-XRT and \textit{HXMT}-LE), and the second one extended with additional optical and near-infrared measurements collected from the literature: see \S\ref{sec:modelling}. This Bayesian Inference was done using the Nuclear physics and Multi-Messenger Astronomy framework, \texttt{NMMA} \citep{Dietrich:2020efo,Pang:2022rzc}, employing the semi-analytic code \texttt{afterglowpy} for afterglow lightcurve modelling \citep{Ryan:2019fhz} and was complemented with an independent Bayesian inference based on the model by \citet{pellouin} to test the impact of more realistic physics for the jet dynamics; this model accounts for both the early coasting phase and for the emission, by including the synchrotron self-Compton emission with a full treatment in Thomson and Klein-Nishina regimes. We started with the simplest version of the standard GRB afterglow model, i.e. a top-hat jet seen exactly on-axis. Both independent pipelines converged to similar solutions, with a best fit that yields poor fits to some of the observations. This analysis confirms a tension between the standard afterglow model and the observed spectral and temporal evolution, as suggested by \citet{2023Laskar}, based on the closure relations. The smoother transitions, rather than sharp breaks, observed in such a detailed model only moderately improve the predicted lightcurves.
In particular, the high frequency lightcurves, especially in X--rays, are not well reproduced, and the late-time radio flux is overpredicted.
In addition, we note that this best-fit yields an unexpectedly large opening angle for such a bright GRB, and a very dense external medium.
We explored several additional effects: free viewing angle, lateral structure of the jet (power-law or Gaussian), early coasting phase, SSC radiation, or an underlying supernova component. None of these models which include more realistic physics leads to better fits. We, therefore, conclude that the modelling of the GRB 221009A afterglow will require going beyond the most standard afterglow model by, for instance, considering more complex jet structure, or external density, or the contribution of the reverse shock, as also suggested by \citet{2023Laskar,2023arXiv230207906O}.

GRB 221009A is an absolutely unique event, representing not just the nearest extremely energetic GRB, but potentially also the most energetic GRB ever detected. These two factors combined make it by far the brightest GRB ever seen, at the very least a once-in-a-lifetime event, more probably even a millenial one. To have such an event occur when we have a fleet of satellites in space able to detect gamma-rays, and the ground- and space-based capabilities to determine the distance and follow up the afterglow evolution in detail, even by amateur astronomers, is fortuitous indeed. It is unlikely that a chance like this will come again in the coming decades or even centuries, making this an event to be remembered through the ages.

\begin{acknowledgements}
The GRANDMA consortium thanks the amateur participants of the Kilonova-Catcher program and observers from GRANDMA.
GRANDMA thanks the paper writing team managed by D.~A.~Kann.
The Kilonova-Catcher program is supported by the IdEx Universit\'e de Paris Cit\'e, ANR-18-IDEX-0001.
The GRANDMA collaboration thanks G. Parent, E. Maris, F.~Bayard, O.~Aguerre and M.~Richmond for their observations.
This project has received financial support from the CNRS through the MITI interdisciplinary programs. We thank Mathias Schultheis for fruitful discussions regarding extinction selection.
D.~A.~Kann acknowledges the support by the State of Hessen within the Research Cluster ELEMENTS (Project ID 500/10.006). 
S.~Antier acknowledges the financial support of the Programme National Hautes Energies (PNHE) and of Cr\'edits Scientifiques Incitatifs d’UCA 2023.
M.~W.~Coughlin acknowledges support from the National Science Foundation with grant numbers PHY-2010970 and OAC-2117997. C.~Andrade and M.~W.~Coughlin were supported by the Preparing for Astrophysics with LSST Program, funded by the Heising Simons Foundation through grant 2021-2975, and administered by Las Cumbres
Observatory.
J.-G. Ducoin is supported by a research grant from the Ile-de-France Region within the framework of the Domaine d’Int\'er\^et Majeur-Astrophysique et Conditions d’Apparition de la Vie (DIM-ACAV). This work has made use of the Infinity Cluster hosted by Institut d'Astrophysique de Paris. 
C. Pellouin acknowledges funding support from the Initiative Physique des Infinis (IPI), a research training program of the Idex SUPER at Sorbonne Universit\'e.
The Egyptian team acknowledges support from the Science, Technology \& Innovation Funding Authority (STDF) under grant number 45779.
S. Karpov is supported by European Structural and Investment Fund and the Czech Ministry of Education, Youth and Sports (Project CoGraDS -- CZ.02.1.01/0.0/0.0/15\_003/0000437).
J.~Mao is supported by the NSFC 11673062 and Oversea Talent Program of Yunnan Province.
The \textit{Insight}-\textit{HXMT} team acknowledges support from the National Key R\&D Program of China (2021YFA0718500) and the National Natural Science Foundation of China under Grants Nos. U1838201, U1838202.
\end{acknowledgements}

\bibliography{references}{}
\bibliographystyle{aasjournal}

\appendix

%\subsection{\textbf{Figures related to the multi-wavelength of the afterglow}}

\begin{figure*}%[!th]
\centering
\includegraphics[width=\textwidth]{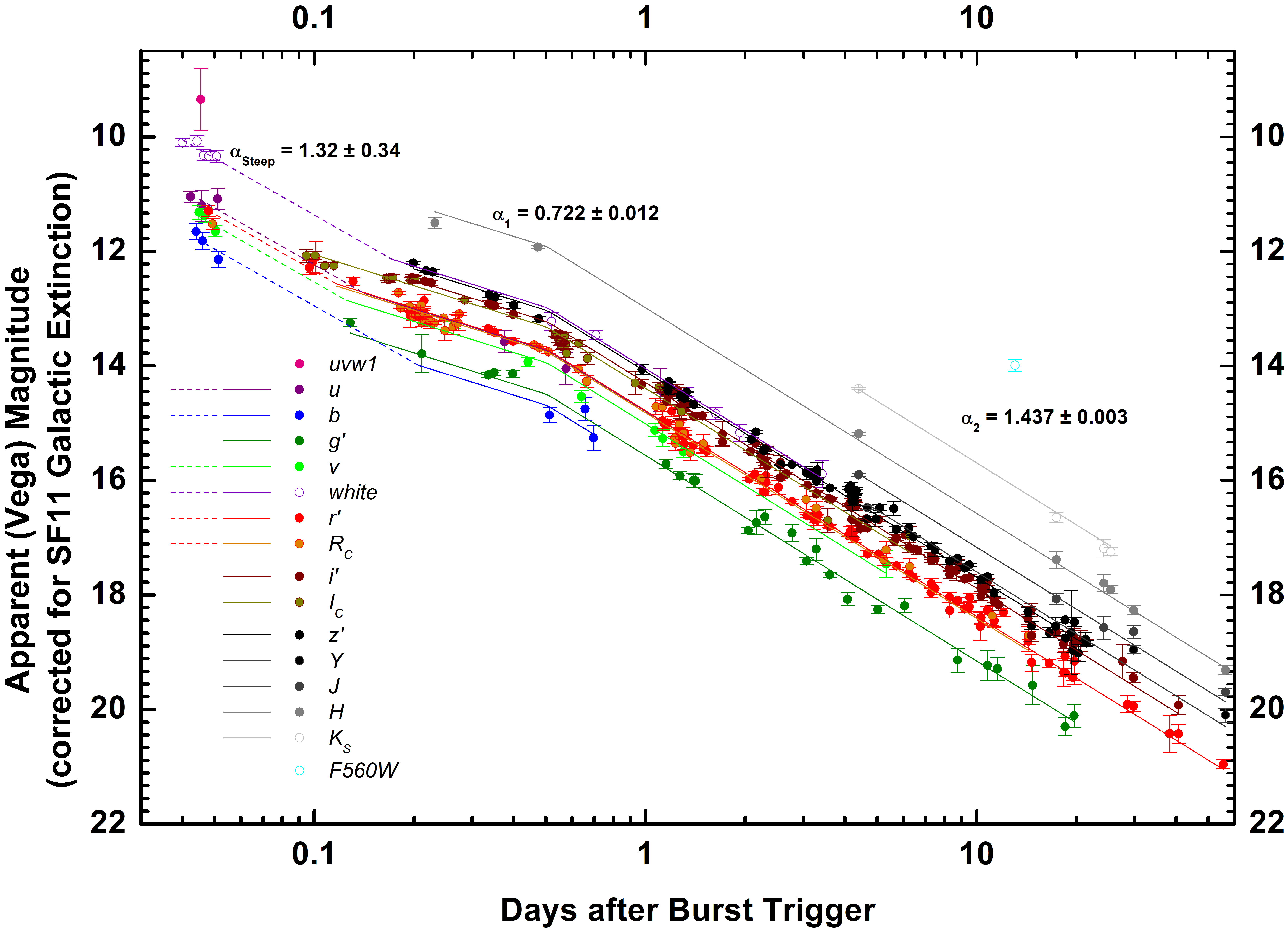}
  \caption{UVOIR light curve of GRB 221009A \textbf{(see section "Empirical light curve analysis)}. The magnitudes, expressed in the Vega system, are corrected for the SF11 galactic extinction. The break slope is at $\sim$ 0.6 d post GRB trigger time between $\alpha_1$ and $\alpha_2$.}
   \label{fig:LC_ALL}
\end{figure*}

\begin{figure}[t]
\centering
\includegraphics[width=0.55\columnwidth]{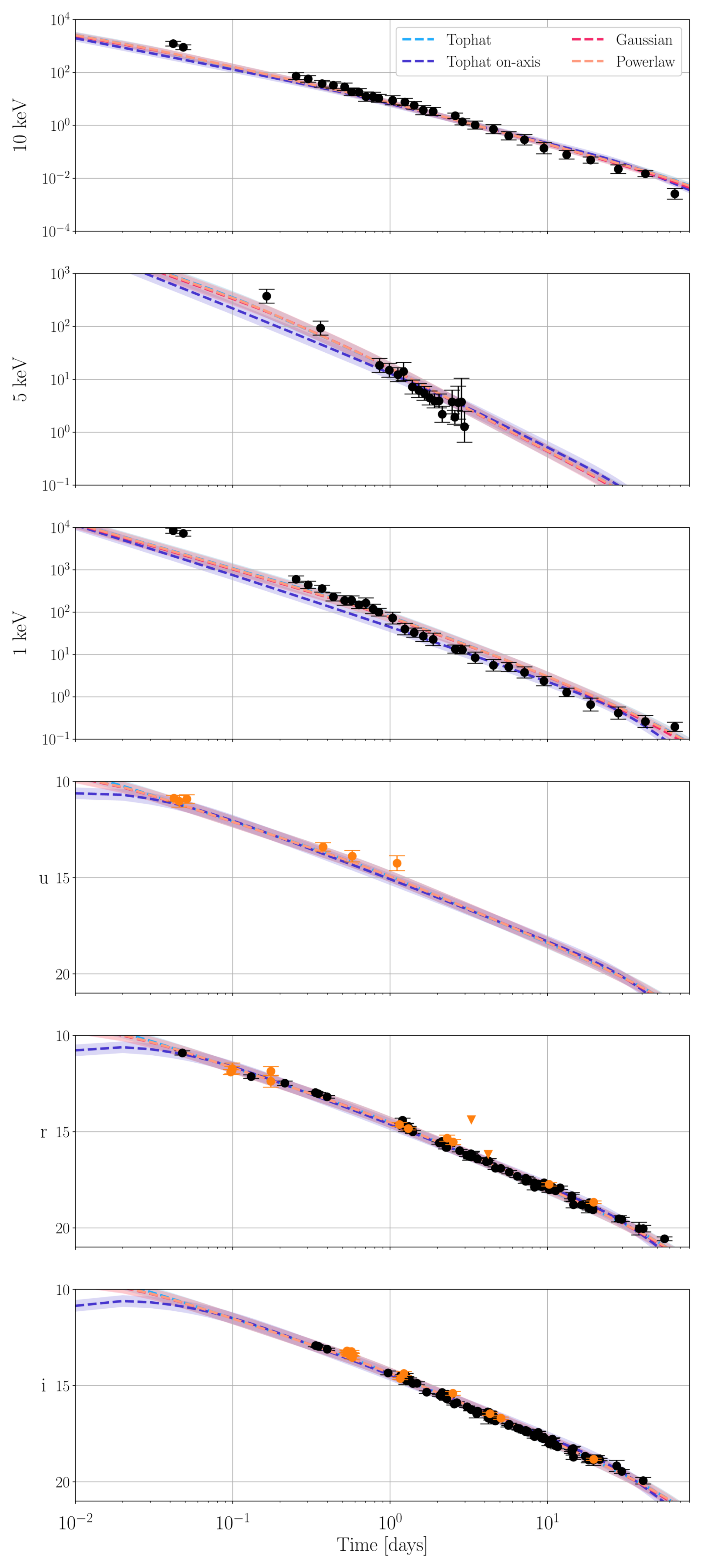}
  \caption{ \texttt{NMMA} - Observational data (\textit{Extended} data) and best-fit light curves of selected filters for the NMMA analysis using the SF11 extinction and the four employed jet structures. The X-ray bands are shown in $\mu$Jy and the rest of the bands are shown in AB magnitude. In the optical band, the \textit{GRANDMA} data points are shown in orange.}
   \label{fig:NMMA_light curve}
\end{figure}

\begin{figure*}[t]
\centering
\includegraphics[width=\textwidth]{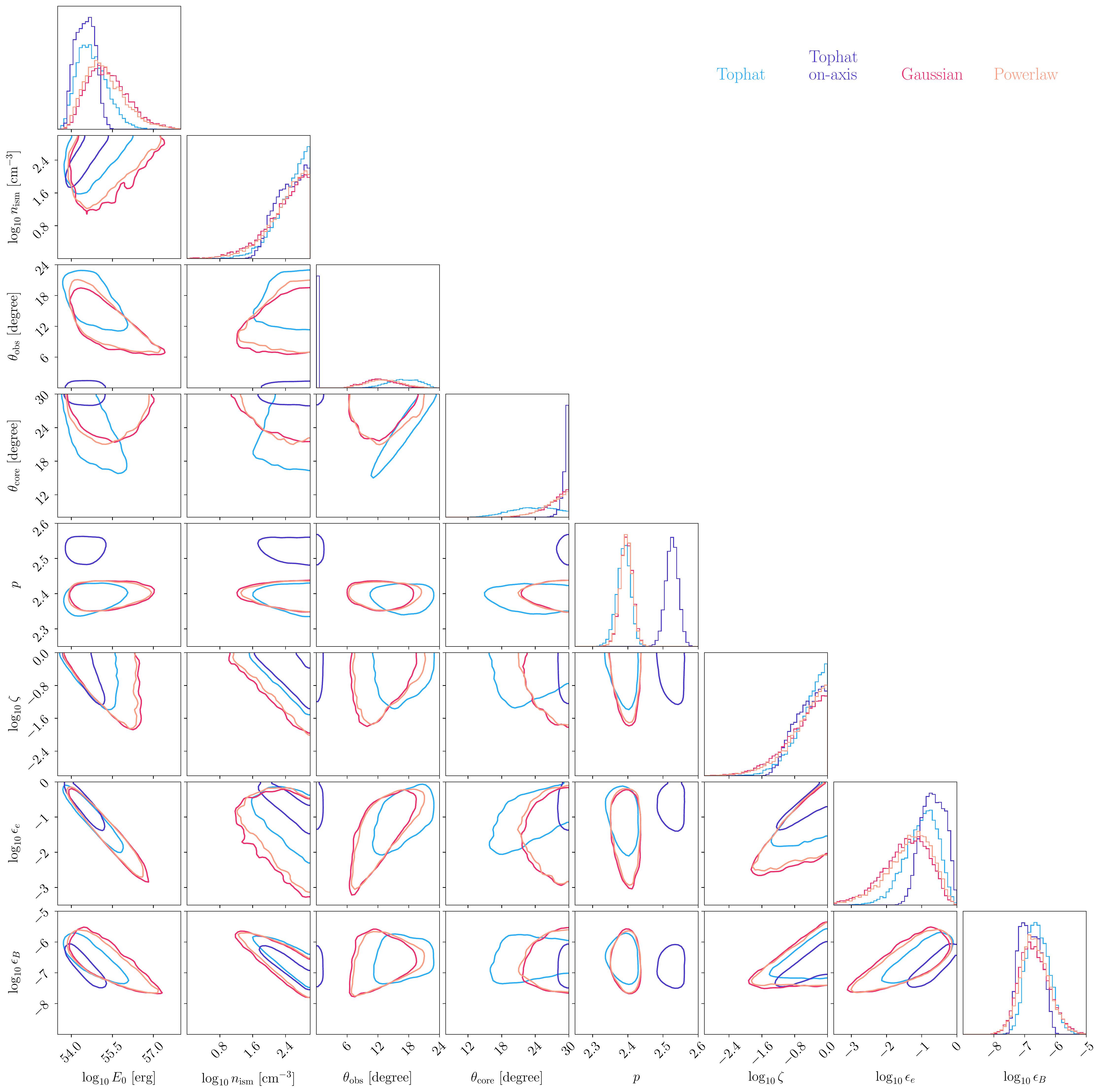}
  \caption{ \texttt{NMMA} - Posterior distribution (shown are 90\% confidence intervals) for our selected data sets when using different jet models of \texttt{afterglowpy} analyzing the \textit{Extended} data.}
  \label{fig:NMMA_corner}
\end{figure*}
\newpage\clearpage
\begin{table*}
\caption{Data used for the numerical data analysis sections (\S\ref{NMMA}, \S\ref{IAP}). \textit{Swift} data have been converted from the Vega system to the AB system. Data are given fully extinction-corrected, for either SF11 MW foreground extinction (described in section~\ref{SED}, or RF09\footnote{This does not include updated measurements provided after Feb. 05. 2022}, and the corresponding SMC extinction in the host galaxy (\S\ref{SED}).}
\label{tab:GRANDMA_observations}
\scalebox{0.9}{
\begin{tabular}{|cc|c|cc|cc|c|}
\hline
\multicolumn{2}{|c|}{Delay} & Filter &  \multicolumn{2}{c|}{SF11}  & \multicolumn{2}{c|}{RF09} & Observatory \\
  (day) & (sec)  &  & Magnitude  & Upper-limit & Magnitude  & Upper-limit & \\
\hline
%\multicolumn{8}{c}{X-ray} \\
%\hline
%0.0487	& $4.209\times10^3$  & $uvw2$ &  - & 9.15 &  - & 14.72 & \textit{Swift} UVOT \\
%\hline
%0.0485	& $4.190\times10^3$ & $uvw1$ & $11.73^{+0.61}_{-0.39}$ & - & $15.86^{+0.61}_{-0.39}$ & -  & \textit{Swift} UVOT \\
%\hline
0.0423 & $3.657\times10^3$ & $u$ & $10.87\pm0.15$ & - & $10.65\pm0.35$ & - & \textit{Swift} UVOT \\	 
0.0457 & $3.949\times10^3$ & $u$ & $11.03\pm0.30$` & - & $10.77\pm0.44$ & -  & \textit{Swift} UVOT \\
0.0511 & $4.416\times10^3$ & $u$ & $10.91\pm0.22$ & - & $10.75\pm0.39$  & - & \textit{Swift} UVOT \\
0.3755 & $3.244\times10^4$ & $u$ & $13.40\pm0.22$ & - & - & - & \textit{Swift} UVOT \\
0.5751 & $4.969\times10^4$ & $u$ & $13.88\pm0.30$ & - & - & - & \textit{Swift} UVOT \\
1.1084 & $9.576\times10^4$ & $u$ & $14.25\pm0.39$ & - & - & - & \textit{Swift} UVOT \\
\hline
\end{tabular}}
\end{table*}

\begin{table*}
\addtocounter{table}{-1}
\caption{Continued.}
\scalebox{0.9}{
\begin{tabular}{|cc|c|cc|cc|c|}
\multicolumn{2}{|c|}{Delay} & Filter &  \multicolumn{2}{c|}{SF11}  & \multicolumn{2}{c|}{RF09} & Observatory \\
  (day) & (sec)  &  & Magnitude  & Upper-limit & Magnitude  & Upper-limit & \\
\hline
2.3042  & $1.991\times10^5$ & $R_C$ & $15.37\pm0.10$ & - & $15.30\pm0.19$ & - & Lisnyky AZT-8  \\
3.2868  & $2.840\times10^5$ & $R_C$ & $16.13\pm0.12$ & - &  $16.06\pm0.20$ & - & Lisnyky AZT-8  \\
5.0674 & $4.378\times10^5$ & $R_C$ & - & 14.61 & - & 14.54 &  UBAI-ST60 \\
11.1313 & $9.617\times10^5$ & $R_C$  & $18.01\pm0.19$ & - & $17.94\pm0.25$ & - & UBAI-AZT22 \\
12.4567 & $1.076\times10^6$ & $R_C$ & - & 16.02 & - & 15.95 & VIRT \\
\hline
0.5333 & $4.608\times10^4$ & $i^\prime$ & $13.20\pm0.06$  & - &  $13.13\pm0.14$  & - &  KNC-BGO \\
\hline
\end{tabular}}
\end{table*}

\begin{table*}
\addtocounter{table}{-1}
\caption{Continued.}
\scalebox{0.9}{
\begin{tabular}{|cc|c|cc|cc|c|}
\multicolumn{2}{|c|}{Delay} & X-ray Filter &  \multicolumn{2}{c|}{Flux}  & \multicolumn{2}{c|}{Flux error} & Observatory \\
  (day) & (sec)  & Central frequency & \multicolumn{2}{c|}{(Jansky)} & \multicolumn{2}{c|}{(Jansky)} & \\
\hline
0.0419 & $0.362\times10^4$  & 10 keV & \multicolumn{2}{c|}{$1.23\times 10^{-3}$} & \multicolumn{2}{c|}{$2.51\times 10^{-4}$} & \textit{Swift} XRT \\
0.0486 & $0.420\times10^4$  & 10 keV & \multicolumn{2}{c|}{$8.95\times 10^{-4}$} & \multicolumn{2}{c|}{$1.88\times 10^{-4}$} & \textit{Swift} XRT \\
0.2529 & $2.185\times10^4$  & 10 keV & \multicolumn{2}{c|}{$7.23\times 10^{-5}$} & \multicolumn{2}{c|}{$2.14\times 10^{-5}$} & \textit{Swift} XRT \\
0.3018 & $2.608\times10^4$  & 10 keV & \multicolumn{2}{c|}{$5.71\times 10^{-5}$} & \multicolumn{2}{c|}{$1.64\times 10^{-5}$} & \textit{Swift} XRT \\
0.3701 & $3.197\times10^4$  & 10 keV & \multicolumn{2}{c|}{$3.76\times 10^{-5}$} & \multicolumn{2}{c|}{$1.10\times 10^{-5}$} & \textit{Swift} XRT \\
0.4365 & $3.771\times10^4$  & 10 keV & \multicolumn{2}{c|}{$3.29\times 10^{-5}$} & \multicolumn{2}{c|}{$9.87\times 10^{-6}$} & \textit{Swift} XRT \\
\hline
\end{tabular}}
\end{table*}

\newpage\clearpage
\begin{center}

\begin{table*}
\caption{The GRANDMA and GCN optical observations of GRB 221009A. In column (2), the $\mathrm{T_{mid}}$ time is the delay between the beginning of the observation and the \textit{Fermi} GBM GRB trigger time (2022-10-09T13:16:59.99) in days. In column (5), magnitudes are given in the AB system and not corrected for Galactic and host-galaxy dust extinction. In column (9), the VETO tag refers to GRANDMA data that were not analyzed due to the bad quality of the images.}
\label{tab:all_observations}
\scalebox{0.6}{
\begin{tabular}{|l|cc|c|l|l|l|c|c|c|}
\hline
\multicolumn{1}{|c|}{T$_{start}$}  & \multicolumn{2}{c|}{T$_{mid}$ (days)}& Filter & Exposure & \multicolumn{1}{c|}{Magnitude} & \multicolumn{1}{c|}{U.L.} & Telescope & Reference & Analysis method \\
\multicolumn{1}{|c|}{UT} & MJD & T-T$_{GRB}$ &  &  &  &  & & & \\
\multicolumn{1}{|c|}{(1)} & \multicolumn{2}{c|}{(2)} & (3) & \multicolumn{1}{|c|}{(4)} & \multicolumn{1}{|c|}{(5)} & \multicolumn{1}{|c|}{(6)} & (7) & (8) & (9) \\
\hline
\hline
2022-10-09T18:06:27	&	59861.7545                	&	0.2010             	&	$u^\prime$ 	&	6$\times$60s       & -	&	17.9 (3$\sigma$)   	&	MeerLICHT      & \cite{2022GCN.32646....1D}   &   -    \\ %ok
2022-10-10T05:00	&	59862.2083                    	&	0.6549 &	$u^\prime$ 	&	6$\times$60s      & -  & -	&	Nickel-1m       & \cite{2022GCN.32669....1V}       &   -  \\ %need to reupdate
\hline
%2022-10-09T19:21	&	59861.8062                    	&	0.2528              	&	$B$     	&	5$\times$180 s & -	&	17.8  	&	\textcolor{black}{KNC-SCT-0.28}   & \cite{2022GCN.32640....1B} \& this work   & APASS DR9 \DT{ from PS1 converted into Johnson B using the transformation in https://ar$\times$iv.org/pdf/2205.06186.pdf}               \\ % Damien tu peu$\times$ refaire l'analyse stp ?
2022-10-09T19:19:51	&	59861.8107       	&	0.2572              	&	$B$     	&	5$\times$180s & -	& 18.1 (5$\sigma$) &	\textcolor{black}{KNC-SCT-0.28}   & \cite{2022GCN.32640....1B} \& this work & \sc{STDpipe}        \\ % UL 17.9 with Muphoten
2022-10-10T04:21	&	59862.1812                    	&	0.6278              	&	$B$     	&	5$\times$60s      	&	-	&	19.7 	&	LOAO-1m     & \cite{2022GCN.32659....1P}   & -             \\ %Ok checked
\hline
2022-10-09T16:21	&	59861.6812                    	&	0.1278              	&	$g^\prime$ 	&	200s       	&	17.66  $\pm$ 0.07  	&	-   	&	GIT        & \cite{2022GCN.32662....1K}    & - \\
\hline
\end{tabular}}
\end{table*}

\begin{table*}
\addtocounter{table}{-1}
\caption{Continued.}
\scalebox{0.6}{
\begin{tabular}{|l|cc|c|l|l|l|c|c|c|}
2022-10-10T12:25	&	59862.5174                    	&	0.9639              	&	$r^\prime$ 	&	2$\times$150s       	&	18.67  $\pm$ 0.16  	&	-   	&	SLT-40      & \cite{2022GCN.32667....1C} &    -         \\
2022-10-10T16:33	&	59862.7023                    	&	1.1489              	&	$r^\prime$ 	&	11$\times$100s   	&	18.65  $\pm$ 0.08  	&	20.9  	&	KAO   & this work  &  \sc{Muphoten} \\
2022-10-10T17:06	&	59862.7125                    	&	1.159               	&	$r^\prime$ 	&	600s       	&	18.65  $\pm$ 0.02  	&	-   	&	RTT-150     & \cite{2022GCN.32743....1B}   & -               \\ % ok checked
2022-10-10T17:14	&	59862.7181                    	&	1.1646              	&	$r^\prime$ 	&	18$\times$30s     	&	18.64  $\pm$ 0.03  	&	20.8 (3$\sigma$)   	&	AZT-20       &  \cite{2022GCN.32670....1K} & -               \\ %ok checked
2022-10-10T19:26	&	59862.8097                    	&	1.2563              	&	$r^\prime$	&	2$\times$300s      	&	18.74  $\pm$ 0.12  	&	-   	&	RC-80         &  \cite{2022GCN.32709....1V} &   -          \\ % ok checked
2022-10-10T19:40	&	59862.8194                    	&	1.266               	&	$r^\prime$ 	&	3$\times$100s     	&	18.8   $\pm$ 0.21  	&	-   	&	LCOGT     & \cite{2022GCN.32693....1S} & -              \\ %ok checked
2022-10-10T20:30	&	59862.8611                    	&	1.3077              	&	$r^\prime$ 	&	2$\times$300s    	&	18.86   $\pm$ 0.04   	&	20.5 	&	\textcolor{black}{C2PU/Omicron}    & this work  & \sc{STDpipe} \& \sc{Muphoten}         \\
2022-10-10T20:50	&	59862.8681                    	&	1.3146              	&	$r^\prime$ 	&	600s       	&	18.81  $\pm$ 0.05  	&	-   	&	RTT-150      & \cite{2022GCN.32743....1B}   & -              \\
\end{tabular}}
\end{table*}

\begin{table*}
\addtocounter{table}{-1}
\caption{Continued.}
\scalebox{0.6}{
\begin{tabular}{|l|cc|c|l|l|l|c|c|c|}
2022-10-13T02:36:45	&	59865.1069                    	&	3.5534              	&	$i^\prime$ 	&	$12\times300$s      	&	19.50  $\pm$ 0.22  	&	19.4 (5$\sigma$)	&	\textcolor{black}{KNC-BGO}     & this work    & \sc{STDpipe}     \\
2022-10-13T02:24:22	&	59865.1149                       	&	3.5614              	&	$i^\prime$ 	&	20$\times$120s      	&	-  	&	18.7 (5$\sigma$)  	&	\textcolor{black}{KNC-iT11}       & this work & \sc{STDpipe}                \\
2022-10-13T03:09	&	59865.1312                    	&	3.5778              	&	$i^\prime$ 	&	-      	&	19.37  $\pm$ 0.05  	&	-   	&	LDT       & \cite{2022GCN.32739....1O} &    -               \\ %ok checked
2022-10-13T17:04	&	59865.7111                    	&	4.1577              	&	$i^\prime$ 	&	600s       	&	19.51  $\pm$ 0.06  	&	-   	&	RTT-150          &  \cite{2022GCN.32752....1B} & -               \\ %ok checked
2022-10-13T17:36	&	59865.7333                    	&	4.1799              	&	$i^\prime$ 	&	600s       	&	19.41  $\pm$ 0.05  	&	-   	&	RTT-150           &  \cite{2022GCN.32752....1B} & -               \\ %ok checked
2022-10-13T18:09	&	59865.7562                    	&	4.2028              	&	$i^\prime$ 	&	600s       	&	19.52  $\pm$ 0.05  	&	-   	&	RTT-150           &  \cite{2022GCN.32752....1B} & -              \\ %ok checked\
2022-10-13T17:42	&	59865.7642                    	&	4.2108              	&	$i^\prime$ 	&	21$\times$110s   	&	19.73  $\pm$ 0.3   	&	20.3  	&	\textcolor{black}{KAO}     & this work & \sc{Muphoten}                 \\
2022-10-13T18:41	&	59865.7785                    	&	4.225               	&	$i^\prime$ 	&	600s       	&	19.44  $\pm$ 0.04  	&	-   	&	RTT-150           &  \cite{2022GCN.32752....1B} & -               \\
\hline
\end{tabular}}
\end{table*}

%2022-10-11T18:14:21 &    59863.75996527778              &   2.2065                  &   I       &   -       &   18.19   &   AbAO    %52$\times$60s

%2022-10-10T15:53:44 &   59862.662314814814              &   1.10885                &    Rc      &   19.27  $\pm$ 0.26   &   18.35   &   ShAO    %15$\times$90 s
%2022-10-10T15:58:41 &   59862.66575231482               &   1.11229                &    V       &   19.85  $\pm$ 0.37   &   19.50   &   ShAO    %25$\times$120 s

\end{center}

\end{document}